\documentclass[11pt]{article}
\usepackage{amssymb}
\usepackage{amsfonts}
\usepackage{amsmath}
\usepackage{geometry}
\usepackage{amsopn}
\usepackage{amstext}
\usepackage{amsxtra}

\newtheorem{theorem}{Theorem}

\newtheorem{claim}{Claim}

\newtheorem{lemma}{Lemma}

\newtheorem{proposition}{Proposition}
\newtheorem{remark}{Remark}

\newenvironment{proof}[1][Proof]{\noindent\textbf{#1.} }{\ \rule{0.5em}{0.5em}}
\input{tcilatex}
\begin{document}

\title{The Empirical Content of Binary Choice Models\thanks{%
Keywords: Binary choice, general heterogeneity, income effect, utility
maximization, integrability/rationalizability, Slutsky inequality,
shape-restrictions. JEL Codes: C14, C25, D12.}}
\author{Debopam Bhattacharya\thanks{%
The author would like to thank the Editor, three anonymous referees, Michael
Floater, Arthur Lewbel, Oliver Linton and seminar participants at several
institutions for helpful feedback. Financial support from the European
Research Council via a Consolidator Grant EDWEL, Project number 681565 is
gratefully acknowledged.} \\
University of Cambridge}
\date{September 17, 2020}
\maketitle

\begin{abstract}
An important goal of empirical demand analysis is choice and welfare
prediction on counterfactual budget sets arising from potential
policy-interventions. Such predictions are more credible when made without
arbitrary functional-form/distributional assumptions, and instead based
solely on economic rationality, i.e. that choice is consistent with utility
maximization by a heterogeneous population. This paper investigates
nonparametric economic rationality in the empirically important context of
binary choice. We show that under general unobserved heterogeneity, economic
rationality is equivalent to a pair of Slutsky-like shape-restrictions on
choice-probability functions. The forms of these restrictions differ from
Slutsky-inequalities for continuous goods. Unlike McFadden-Richter's
stochastic revealed preference, our shape-restrictions (a) are global, i.e.
their forms do not depend on which and how many budget-sets are observed,
(b) are closed-form, hence easy to impose on parametric/semi/non-parametric
models in practical applications, and (c) provide computationally simple,
theory-consistent bounds on demand and welfare predictions on counterfactual
budget-sets.
\end{abstract}

\section{Introduction}

Many important economic decisions faced by individuals are binary in nature,
including labour force participation, retirement, college enrolment,
adoption of a new technology or health product, participation in a
job-training program, etc. This paper concerns nonparametric analysis of
binary choice under general unobserved heterogeneity and income effects. The
paper has two goals. The first is to understand, theoretically, what
nonparametric restrictions does utility maximization by heterogeneous
consumers impose upon choice-probabilities, i.e. whether there are analogs
of Slutsky restrictions for binary choice under general unobserved
heterogeneity and income effects, and conversely, whether these restrictions
are also sufficient for observed choice-probabilities to be rationalizable.
This issue is important for logical coherency between theory and empirics
and for prediction of demand and welfare in situations involving
counterfactual, i.e. previously unobserved, budget sets. It is important in
these exercises to allow for general unobserved heterogeneity because
economic theory typically does not restrict its dimension or distribution,
and does not specify how it enters utility functions. To date, closed-form
Slutsky conditions for rationalizability of demand under general
heterogeneity were available only for continuous choice. The present paper,
to our knowledge, is the first to establish them for the leading case of
discrete demand, viz. binary choice.

The second goal of the present paper is a practical one. It is motivated by
the fact that in empirical applications of binary choice, requiring the
estimation of elasticities, welfare calculations and demand predictions,
researchers typically use parsimonious functional-forms for conditional
choice probabilities. This is because fully nonparametric estimation is
often hindered by curse of dimensionality, the sensitivity of estimates to
the choice of tuning parameters and insufficient price variation, especially
in consumer data from developed countries. The question therefore arises as
to whether the economic theory of consumer behavior can inform the choice of
such functional forms. Answering this question is our second objective.

Since McFadden 1973, discrete choice models of economic behavior have been
studied extensively in the econometric literature, mostly under restrictive
assumptions on utility functions and unobserved heterogeneity including,
inter alia, quasi-linear preferences implying absence of income effects
and/or parametrically specified heterogeneity distributions (c.f. Train 2009
for a textbook treatment). Matzkin (1992) investigated the nonparametric
identification of binary choice models with additive heterogeneity, where
both the distribution of unobserved heterogeneity and the functional form of
utilities were left unspecified. More recently, Bhattacharya (2015, 2018)
has shown that in discrete choice settings, welfare distributions resulting
from price changes are nonparametrically point-identified from choice
probabilities without any substantive restriction on preference
heterogeneity, and even when preference distribution and heterogeneity
dimension are not identified.

In the present paper, we consider a setting of binary choice by a population
of budget-constrained consumers with general, unobserved heterogeneity,
producing an individual-level cross-sectional dataset that records prices,
individual income and the choice made by the individual.\footnote{%
As a referee has correctly commented, income plays a prominent role in this
paper, unlike many existing empirical applications which ignore the role of
income.} In this setting, we develop a characterization of utility
maximization which takes the form of simple, closed-form shape restrictions
on choice probability functions in the population. These nonparametric
shape-restrictions can be consistently tested in the usual asymptotic
econometric sense and are extremely easy to impose on specifications of
choice-probabilities -- akin to testing or imposing monotonicity of
regression functions. Most importantly, they lead to computationally simple
bounds for theory-consistent demand and welfare predictions on
counterfactual budgets sets -- an important goal of empirical demand
analysis. Interestingly, our shape-restrictions differ in form from the
well-known Slutsky inequalities for continuous goods.

The above results are developed in a fully nonparametric context;
nonetheless, they can help guide applied researchers intending to use simple
parametric or semiparametric models. As a specific example, consider the
popular probit/logit type model for binary choice of whether to buy a
product or not.\ A standard specification is that the probability of buying
depends (implicitly conditioning on other observed covariates) on its price $%
p$ and the decision-maker's income $y$, e.g. $\bar{q}\left( p,y\right)
=F\left( \gamma _{0}+\gamma _{1}p+\gamma _{2}y\right) $, where $F\left(
\cdot \right) $ is a distribution function. We will show below that these
choice-probabilities are consistent with utility maximization by a
heterogenous population of consumers, if and only if $\gamma _{1}\leq 0$,
and $\gamma _{1}+\gamma _{2}\leq 0$. While the first inequality simply means
that demand falls with own price (holding income fixed), the second
inequality is less obvious, and constitutes an important empirical
characterization of utility maximization.

For the case of \textit{continuous} goods, Lewbel 2001 explored the question
of when average demand, generated from maximization of heterogeneous
individual preferences, satisfies standard properties of non-stochastic
demand functions. More recently, for the case of two continuous goods (i.e.
a good of interest plus the numeraire) under general heterogeneity, Dette,
Hoderlein and Neumayer 2016 have shown that constrained utility maximization
implies quantiles of demand satisfy standard Slutsky negativity, and Hausman
and Newey 2016 have shown that the two are in fact equivalent. The analog of
the two goods setting in discrete choice is the case of binary alternatives.
Accordingly, our main result (Theorem 1 below) may be viewed as the discrete
choice counterpart of Hausman and Newey 2016, Theorem 1. Note however that
quantiles are degenerate for binary outcomes, and indeed, the forms of our
Slutsky-like shape restrictions are completely different from Dette et al
and Hausman-Newey's quantile-based conditions for continuous choice.

An alternative, algorithmic -- as opposed to closed-form and analytic --
approach to rationalizability of demand is the \textquotedblleft revealed
stochastic preference\textquotedblright\ (SRP, henceforth) method, which
applies to very general choice settings where a heterogeneous population of
consumers faces a finite number of budget sets, c.f. McFadden and Richter
1990, McFadden 2005. When budget sets are numerous or continuously
distributed, as in household surveys with many income and/price values, SRP
is well-known to be operationally prohibitive, c.f. Anderson et al 1992,
Page 54-5 and Kitamura and Stoye 2016, Sec 3.3. Furthermore, the SRP
conditions are difficult to impose on parametric specifications commonly
used in practical applications, they change entirely in form upon addition
of new budget sets, and are cumbersome to use for demand prediction on
counterfactual budgets, especially in welfare calculations that typically
require simultaneous prediction of demand on a continuous range of
budget-sets. In contrast, our approach yields rationality conditions which
(a) are global, in that they characterize choice probability \textit{%
functions}, and their forms remain invariant to which and how many budget
sets are observed in a dataset, and (b) are closed-form, analytic
shape-restrictions, hence easy to impose, standard to test, and simple to
use for the important practical problem of counterfactual predictions of
demand and welfare. As such, these shape-restrictions establish the analogs
of Slutsky conditions -- the cornerstone of classical demand analysis -- for
binary choice under general unobserved heterogeneity and income effects.

\section{The Result}

Consider a population of heterogeneous individuals, each choosing whether or
not to buy an indivisible good. Let $N$ represent the quantity of numeraire
which an individual consumes in addition to the binary good. If the
individual has income $Y=y$, and faces a price $P=p$ for the indivisible
good, then the budget constraint is $N+pQ=y$ where $Q\in \left\{ 0,1\right\} 
$ represents the binary choice. Individuals derive satisfaction from both
the indivisible good as well as the numeraire. Upon buying, an individual
derives utility from the good but has a lower amount of numeraire $y-p$
left; upon not buying, she enjoys utility from her outside option and a
higher quantity of numeraire $y$. There is unobserved heterogeneity across
consumers which affect their choice, and so on each budget set defined by a
price $p$ and consumer income $y$, there is a (structural) probability of
buying, denoted by $\bar{q}\left( p,y\right) $; that is, if each member of
the entire population were offered income $y$ and price $p$, then a fraction 
$\bar{q}\left( p,y\right) $ would buy the good. For now, we implicitly
condition our analysis on observed covariates, and later show how to
incorporate them into the results. We will show that these choice
probabilities will be consistent with utility maximization by a
heterogeneous population if and only if the following Slutsky-like conditions%
\footnote{%
Our main result does not need smoothness; we write the conditions with
derivatives here to show the Slutsky-like form of the result.} hold:%
\begin{equation}
\frac{\partial }{\partial p}\bar{q}\left( p,y\right) \leq 0\text{, and }%
\frac{\partial }{\partial p}\bar{q}\left( p,y\right) +\frac{\partial }{%
\partial y}\bar{q}\left( p,y\right) \leq 0\text{.}  \label{11}
\end{equation}

For establishing this result, it will be convenient to rewrite the choice
probabilities in an equivalent way as $q\left( y,y-p\right) =\bar{q}\left(
p,y\right) $. Indeed, one can go back and forth between the two
specifications because $\bar{q}\left( c,d\right) \equiv q\left( d,d-c\right) 
$ and $q\left( a,b\right) \equiv \bar{q}\left( a-b,a\right) $. The $q\left(
y,y-p\right) $ formulation is motivated by the fact that given the budget
set $\left( P,Y\right) =\left( p,y\right) $, an individual faces choice
between the bundles $\left( 0,y\right) $ and $\left( 1,y-p\right) $; thus $%
q\left( \cdot ,\cdot \right) $ is an equivalent representation of choice
probabilities as functions of the income left over upon choosing options 0
and 1, respectively. For ease of exposition, we will state our results in
terms of $q\left( \cdot ,\cdot \right) $, and show that under smoothness
they reduce to restriction (\ref{11}) on $\bar{q}\left( \cdot ,\cdot \right) 
$.

The following theorem establishes conditions that are necessary and
sufficient for the conditional choice probability function to be generated
from utility maximization by a heterogeneous population, where no a priori
restriction is imposed on the dimension and functional form of the
distribution of unobserved heterogeneity or on the functional form of
utilities.

To formally state the theorem, we introduce some notation. Let $\bar{\Omega}$
denote the support of $\left( P,Y\right) $; let $\Omega _{1}=\left\{
y-p:\left( p,y\right) \in \bar{\Omega}\right\} $ denote the support of $Y-P$%
, and for any $a_{1}\in \Omega _{1}$ let $\Omega _{0}\left( a_{1}\right)
=\left\{ y:\left( p,y\right) \in \bar{\Omega}\text{, }y-p=a_{1}\right\} $.
Corresponding to the support $\bar{\Omega}$ of $\left( P,Y\right) $, denote
the support of $\left( Y,Y-P\right) $ by $\Omega $, as short-hand for $\cup
_{a_{1}\in \Omega _{1}}\cup _{a_{0}\in \Omega _{0}\left( a_{1}\right)
}\left\{ a_{0},a_{1}\right\} $.\medskip

\begin{theorem}
For binary choice under general heterogeneity, the following two statements
are equivalent:

(I) The structural choice probability function $q\left( \cdot ,\cdot \right)
:\Omega \rightarrow \left[ 0,1\right] $ satisfies that (A) (i) $q\left(
\cdot ,y-p\right) $ is non-increasing, and (ii) $q\left( y,\cdot \right) $
is non-decreasing; (B) $q\left( \cdot ,y-p\right) $ is continuous; (C)
corresponding to any fixed value $a_{1}\in \Omega _{1}$, there exist a small
enough real number $y_{L}\left( a_{1}\right) \in \Omega _{0}\left(
a_{1}\right) $, satisfying $\lim_{y\searrow y_{L}\left( a_{1}\right)
,y-p=a_{1}}q\left( y,y-p\right) =1$ and a large enough real number $%
y_{H}\left( a_{1}\right) \in \Omega _{0}\left( a_{1}\right) $, satisfying $%
\lim_{y\nearrow y_{H}\left( a_{1}\right) ,y-p=a_{1}}q\left( y,y-p\right) =0$.

(II) There exists a pair of utility functions $W_{0}\left( \cdot ,\eta
\right) $ and $W_{1}\left( \cdot ,\eta \right) $, where the first argument
denotes the amount of numeraire, and $\eta $ denotes unobserved
heterogeneity, and a distribution $G\left( \cdot \right) $ of $\eta $ such
that%
\begin{equation*}
q\left( y,y-p\right) =\int 1\left\{ W_{0}\left( y,\eta \right) \leq
W_{1}\left( y-p,\eta \right) \right\} dG\left( \eta \right) \text{,}
\end{equation*}%
where (A') for each fixed $\eta $, (i) $W_{0}\left( \cdot ,\eta \right) $ is
continuous and strictly increasing, and (ii) $W_{1}\left( \cdot ,\eta
\right) $ is non-decreasing; (B') for any $p,y\in \bar{\Omega}$, it holds
that $\int 1\left\{ W_{1}\left( y-p,\eta \right) =W_{0}\left( y,\eta \right)
\right\} dG\left( \eta \right) =0$; (C') corresponding to any fixed $%
a_{1}\in \Omega _{1}$, there exist a small enough real number $y_{L}\left(
a_{1}\right) \in \Omega _{0}\left( a_{1}\right) $ and a large enough real
number $y_{H}\left( a_{1}\right) \in \Omega _{0}\left( a_{1}\right) $,
satisfying $\lim_{y\searrow y_{L}\left( a_{1}\right) ,y-p=a_{1}}\Pr \left[
W_{0}\left( y,\eta \right) \leq W_{1}\left( y-p,\eta \right) \right] =1$ and 
$\lim_{y\nearrow y_{H}\left( a_{1}\right) ,y-p=a_{1}}\Pr \left[ W_{0}\left(
y,\eta \right) \leq W_{1}\left( y-p,\eta \right) \right] =0$.
\end{theorem}

\begin{proof}
In Appendix
\end{proof}

The key step in the proof is showing that (I) implies (II). This is done by
constructing the utility functions $W_{0}\left( y,\eta \right) =y$ and $%
W_{1}\left( y-p,\eta \right) =q^{-1}\left( V,y-p\right) $ with $q^{-1}\left(
\cdot ,y-p\right) $ denoting a suitably defined inverse of the function $%
q\left( \cdot ,y-p\right) $ with respect to its first argument, and the
random variable $\eta =V\sim Uniform(0,1)$. Under conditions A, B,\ C of
Theorem 1, this construction is then shown to imply that $\Pr \left[
W_{1}\left( y-p,\eta \right) \geq W_{0}\left( y,\eta \right) \right]
=q\left( y,y-p\right) $. The formal proof appears in the Appendix.

\textbf{Interpretation of conditions}: Intuitively, conditions (A/A') mean
that having more numeraire ceteris paribus is (weakly) better for every
consumer, i.e. preferences are increasing in the amount of income left over
after any choice. Condition (B/B') -- the \textquotedblleft
no-tie\textquotedblright\ assumption -- is standard in discrete choice
models, and intuitively means that there is a continuum of tastes. Condition
(C) adds to condition (A); it says that holding fixed the income left over
upon choosing option 1, if the income left over upon choosing option 0 is,
hypothetically\textit{,} made small enough, then everyone, i.e. all $\eta $,
will choose option 1. In particular, $y\searrow y_{L}\left( a_{1}\right)
,y-p=a_{1}$ means that starting from a situation with $y-p=a_{1}$, we are
lowering $p$ and $y$ by equal amounts, keeping $y-p$, i.e. the income left
over upon choosing option 1, fixed at $a_{1}$ while $y$, the income left
over upon choosing option 0, is lowered toward $y_{L}\left( a_{1}\right) $,
i.e., $q\left( \underset{\searrow y_{L}\left( a_{1}\right) }{\underbrace{y}},%
\underset{\text{fixed at }a_{1}}{\underbrace{y-p}}\right) \nearrow 1$. A
symmetric interpretation applies to $y_{H}\left( a_{1}\right) $. The
following examples illustrate Condition C.

\textbf{Example 1} (High and Low Price): Suppose $0,1$ denote respectively
not buying and buying a binary good. Suppose preferences are such that at
any income $y$, if price takes a high enough value $p^{H}$, e.g. close to
the highest income in the population, no one would buy the good; conversely,
when price takes a low enough value $p^{L}$, e.g. the good is free ($p^{L}=0$%
) or there is a high enough reward $r>0$ for choosing option 1 (i.e. $%
p^{L}=-r<0$) as in conditional cash transfer programs for school-attendance,
everyone (i.e. all $\eta $) will choose option 1. Then starting from $%
y-p=a_{1}>0$, raising $p$ towards $p^{H}$ while simultaneously increasing $y$
by equal amount keeping $y-p$, the income left upon buying, fixed at $a_{1}$%
, we have that $q\left( y,y-p\right) \equiv q\left( a_{1}+p,a_{1}\right)
\searrow q\left( a_{1}+p^{H},a_{1}\right) =0$; similarly, letting $p\searrow
p^{L}$ and $y\searrow a_{1}+p^{L}$ while keeping fixed $y-p=a_{1}>0$, we
have that $q\left( \underset{\searrow a_{1}+p^{L}}{\underbrace{y}},\underset{%
\text{fixed at }a_{1}}{\underbrace{y-p}}\right) \nearrow q\left(
a_{1}+p^{L},a_{1}\right) =1$. Thus $y_{H}\left( a_{1}\right) =a_{1}+p^{H}$,
and $y_{L}\left( a_{1}\right) =a_{1}+p^{L}$.

\textbf{Example 2 }(Labour supply): Suppose $0,1$ denote not working and
working, respectively, $y$ is non-labour income (e.g. spousal earning or
interest income from investment), and $p=-w$ is the negative of net wage
received upon working, so that $q\left( y,y-p\right) =q\left( y,y+w\right) $%
. Here it is natural to assume that if non-labour income $y$ is zero, then
an individual must work at any positive net wage $w$ for subsistence, so
that $q\left( 0,w\right) =1$, and thus $y_{L}\left( a_{1}\right) =0$ for any
positive $a_{1}$. Similarly, if net wage is zero, then no one with positive
non-labour income will work, i.e. $q\left( y,y\right) =0$, and thus $%
y_{H}\left( a_{1}\right) =a_{1}$.

\begin{remark}
Condition C/C', which simplify the proof of the Theorem, can be dropped. In
the appendix, we provide an alternative version of the theorem without
conditions (C/C'), but with a slightly stronger continuity requirement
(B/B') and a significantly longer proof.
\end{remark}

\begin{remark}
Note that assumptions (A)-(C) place \textit{no restriction} on income
effects, including its sign.
\end{remark}

In statement (II) in Theorem 1, the functions $W_{j}\left( x,\eta \right) $
will correspond to the utility from choosing alternative $j\in \left\{
0,1\right\} $ and being left with a quantity $x$ of the numeraire, and with $%
\eta $ denoting unobserved heterogeneity. This notation allows for the case
where different vectors of unobservables enter the two utilities, i.e. where
the utilities are given by $u_{0}\left( \cdot ,\eta _{0}\right) $ and $%
u_{1}\left( \cdot ,\eta _{1}\right) $, respectively, with $\eta _{0}\neq
\eta _{1}$; simply set $\eta \equiv \left( \eta _{0},\eta _{1}\right) $, $%
W_{0}\left( \cdot ,\eta \right) \equiv u_{0}\left( \cdot ,\eta _{0}\right) $%
, $W_{1}\left( \cdot ,\eta \right) \equiv u_{1}\left( \cdot ,\eta
_{1}\right) $. In the proof of the above theorem, when showing (II) implies
(I), $\eta $ will be allowed to have \textit{any arbitrary and unknown }%
dimension and distribution; in showing (I) implies (II) we will construct a
scalar heterogeneity distribution that will rationalize the choice
probabilities (see further discussion on this point under the heading
"Observational Equivalence" in the next section).

\section{Further Discussion}

\textbf{A. Slutsky Form}: To see the analogy between the shape restrictions
in Theorem 1 and the traditional Slutsky inequality constraints with smooth
demand, rewrite the choice probability on a budget set $\left( p,y\right) $
in the standard form as a function of price and income, viz. $\bar{q}\left(
p,y\right) \equiv q\left( y,y-p\right) $ i.e., $q\left( a_{0},a_{1}\right)
\equiv \bar{q}\left( a_{0}-a_{1},a_{0}\right) $. Then, under continuous
differentiability, the shape restrictions (A) from Theorem 1 are equivalent
to%
\begin{eqnarray}
\frac{\partial }{\partial p}\bar{q}\left( p,y\right) &=&\left. -\frac{%
\partial q\left( a_{0},a_{1}\right) }{\partial a_{1}}\right\vert
_{a_{0}=y,a_{1}=y-p}\leq 0\text{, by Thm 1, (Aii)}  \label{3} \\
\frac{\partial }{\partial p}\bar{q}\left( p,y\right) +\frac{\partial }{%
\partial y}\bar{q}\left( p,y\right) &=&\left. -\frac{\partial q\left(
a_{0},a_{1}\right) }{\partial a_{1}}+\frac{\partial q\left(
a_{0},a_{1}\right) }{\partial a_{0}}+\frac{\partial q\left(
a_{0},a_{1}\right) }{\partial a_{1}}\right\vert _{a_{0}=y,a_{1}=y-p}  \notag
\\
&=&\left. \frac{\partial q\left( a_{0},a_{1}\right) }{\partial a_{0}}%
\right\vert _{a_{0}=y,a_{1}=y-p}\leq 0\text{, by Thm 1, (Ai)}  \label{2}
\end{eqnarray}%
for all $p,y$.\footnote{%
I am grateful to a referee for suggesting this way of showing the
equivalence.} The forms of these inequalities are distinct from textbook
Slutsky conditions for \textit{nonstochastic} demand $q^{\ast }\left(
p,y\right) $ for a \textit{continuous} good, which are given by%
\begin{equation}
\frac{\partial }{\partial p}q^{\ast }\left( p,y\right) +q^{\ast }\left(
p,y\right) \frac{\partial }{\partial y}q^{\ast }\left( p,y\right) \leq 0%
\text{ for all }p,y\text{.}  \label{15}
\end{equation}%
For a continuous good and under general unobserved heterogeneity, Dette,
Hoderlein and Neumeyer 2016 (building on earlier work of Hoderlein 2011),
and Hausman and Newey 2016 show that (\ref{15}) also holds with $q^{\ast
}\left( p,y\right) $ denoting any quantile of the demand distribution for
fixed $\left( p,y\right) $. Thus, for binary choice with general
heterogeneity, the forms of the Slutsky inequality (\ref{3}) and (\ref{2})
are different from the continuous choice counterpart (\ref{15}).\footnote{%
Bhattacharya, 2015 (see also Lee and Bhattacharya, 2018) noted that (\ref{3}%
) (resp, (\ref{2})) is necessary for the CDF of equivalent variation (resp.,
compensating variation) resulting from price-changes to be non-decreasing.}
In particular, the inequalities (\ref{3}) and (\ref{2}) are \textit{linear}
in $\bar{q}\left( \cdot ,\cdot \right) $ (and $q\left( \cdot ,\cdot \right) $%
), unlike (\ref{15}), and hence easier to impose on nonparametric estimates
of $q\left( \cdot ,\cdot \right) $ using, say, shape-preserving sieves that
guarantee that $\frac{\partial }{\partial a_{1}}\hat{q}\left(
a_{0},a_{1}\right) \geq 0$, and $\frac{\partial }{\partial a_{0}}\hat{q}%
\left( a_{0},a_{1}\right) \leq 0$ for all $a_{0},a_{1}$.

\begin{remark}
It is tempting to think of (\ref{3}) and (\ref{2}) as (\ref{15}) with the
level $q^{\ast }\left( p,y\right) $ replaced by 0 and 1 corresponding to
either of the two possible individual choices. However, this interpretation
is incorrect, since $\bar{q}\left( p,y\right) $ is \textbf{average} demand,
and takes values strictly inside $\left( 0,1\right) $. In other words, $\bar{%
q}\left( p,y\right) $ is neither a quantile, nor individual demand at price $%
p$ and $y$, and generically (e.g. in a probit model) does not take the
values of 0 and 1. Thus (\ref{3}) and (\ref{2}) \textbf{cannot} be rewritten
as%
\begin{equation*}
\frac{\partial }{\partial p}\bar{q}\left( p,y\right) +\bar{q}\left(
p,y\right) \frac{\partial }{\partial y}\bar{q}\left( p,y\right) \leq 0\text{
for all }p,y\text{,}
\end{equation*}%
and, as such, are different from the continuous choice counterpart (\ref{15}%
).
\end{remark}

\begin{remark}
Our rationality conditions (A) take the form of simple monotonicity
restrictions on the regression function $q\left( \cdot ,\cdot \right) $.
There are several papers in the Statistics literature on testing
monotonicity of nonparametrically estimated regressions, e.g. Ghosal et al
2000, Hall and Heckman 2000, Chetverikov 2012, etc. which can therefore be
used here.
\end{remark}

\textbf{B. Observational Equivalence}: The construction in our proof of (II) 
$\Rightarrow $ (I) shows that a rationalizable binary choice model with
general heterogeneity of unspecified dimension is observationally equivalent
to one where a scalar heterogeneity enters the utility function of one of
the alternatives in a monotonic way, and the utility of the other
alternative is non-stochastic.\footnote{%
For quantile demand in the continuous case, a result of similar spirit is
discussed in Hausman-Newey, 2016, Page 1228-9, following Theorem 1. In
general, a result holding for the continuous case with two goods does not
necessarily imply that it also holds for the binary case. For example,
welfare related results are different for the binary and the two-good
continuous case, c.f. Hausman-Newey 2016, and Bhattacharya 2015, and so are
Slutsky negativity conditions, as discussed above.} An intuitive explanation
of this equivalence is that in the binary case, choice probabilities are
determined solely by the marginal distribution of reservation price (given
income) for alternative 1, and not the relative ranking of individual
consumers in terms of their preferences within that distribution. So, as
income varies, choice probabilities change only insofar as the marginal
distribution of the reservation price changes, irrespective of how
individual consumers' relative positions change within that distribution.

It is worth pointing out here that a binary choice model with \textit{%
additive} scalar heterogeneity -- the so-called ARUM model -- is
restrictive, and \textit{not} observationally equivalent to a binary choice
model with general heterogeneity. To see this, suppose choice probabilities
are generated via the ARUM model, viz.%
\begin{eqnarray}
q\left( a_{0},a_{1}\right) &=&\Pr \left[ W_{1}\left( a_{1}\right) +\eta
_{1}>W_{0}\left( a_{0}\right) +\eta _{0}\right]  \notag \\
&=&\Pr \left[ \eta _{0}-\eta _{1}<W_{1}\left( a_{1}\right) -W_{0}\left(
a_{0}\right) \right]  \notag \\
&=&F_{\eta _{0}-\eta _{1}}\left[ W_{1}\left( a_{1}\right) -W_{0}\left(
a_{0}\right) \right] \text{.}  \label{7}
\end{eqnarray}%
Assuming smoothness and strict monotonicity of $F_{\eta _{0}-\eta _{1}}\left[
\cdot \right] $, $W_{1}\left( \cdot \right) $ and $W_{0}\left( \cdot \right) 
$, and thus of $q\left( \cdot ,\cdot \right) $, it follows that%
\begin{eqnarray*}
&&\frac{\partial ^{2}}{\partial a_{0}\partial a_{1}}\ln \left[ -\frac{\frac{%
\partial }{\partial a_{1}}q\left( a_{0},a_{1}\right) }{\frac{\partial }{%
\partial a_{0}}q\left( a_{0},a_{1}\right) }\right] \\
&=&\frac{\partial ^{2}}{\partial a_{0}\partial a_{1}}\ln \left( \frac{%
W_{1}^{\prime }\left( a_{1}\right) }{W_{0}^{\prime }\left( a_{0}\right) }%
\right) \text{, from (\ref{7})} \\
&=&\frac{\partial ^{2}}{\partial a_{0}\partial a_{1}}\left[ \ln \left(
W_{1}^{\prime }\left( a_{1}\right) \right) -\ln \left( W_{0}^{\prime }\left(
a_{0}\right) \right) \right] \\
&=&0\text{,}
\end{eqnarray*}%
for every $a_{0}$ and $a_{1}$. This equality is obviously not true for a
general smooth and strictly monotone $q\left( \cdot ,\cdot \right) $
satisfying conditions (A)-(C) of Theorem 1.

\begin{remark}
The construction of $q^{-1}\left( V,\cdot \right) $ in our proof of (II) $%
\Rightarrow $ (I) is unrelated to the almost sure representation of a
continuous random variable $X$ as $F_{X}^{-1}\left( U\right) $ with $%
U=F_{X}\left( X\right) $, where $F_{X}$ and $F_{X}^{-1}$ denote the CDF and
quantile function of $X$, and $U$ is distributed $U\left( 0,1\right) $.
Indeed, if we were to apply this so-called "probability-integral transform"
to $X=W_{1}\left( a_{1},\eta \right) $ for a fixed $a_{1}$, we will have $%
W_{1}\left( a_{1},\eta \right) \overset{a.s.}{=}F_{W_{1}\left( a_{1},\eta
\right) }^{-1}\left( U\left( a_{1}\right) \right) $, where the scalar-valued
uniform process $U\left( a_{1}\right) \equiv F_{W_{1}\left( a_{1},\eta
\right) }\left( W_{1}\left( a_{1},\eta \right) \right) $ will vary with $%
a_{1}$, unlike $V$ in the proof of our theorem above, and therefore cannot
represent unobserved heterogeneity in consumer preferences. In other words,
our constructed $q^{-1}\left( V,a_{1}\right) $ will \textit{not} equal the
data generating process $W_{1}\left( a_{1},\eta \right) $ almost surely, but
the probability that $q^{-1}\left( V,a_{1}\right) \geq a_{0}$ will equal the
probability that $W_{1}\left( a_{1},\eta \right) \geq W_{0}\left( a_{0},\eta
\right) $ for all $\left( a_{0},a_{1}\right) $.\medskip
\end{remark}

\textbf{C. Giffen Goods}: Our rationalizability condition (\ref{3}) says
that own price effect on average demand is negative. This condition has no
counterpart in the continuous case, appears to rule out Giffen behavior and
may, therefore, appear restrictive. We now show that that is not the case:
indeed, Giffen goods cannot arise in binary choice models if utilities are
non-satiated in the numeraire. To see this, let the utility of options $0$
and $1$ be given by $W_{0}\left( \cdot ,\eta \right) $ and $W_{1}\left(
\cdot ,\eta \right) $ as in Theorem 1 above. Now note that if option 1 is
Giffen for an $\eta $ type consumer with income $y$, then for some prices $%
p<p^{\prime }$ she buys at price $p^{\prime }$ but does not buy at $p$.
Therefore,%
\begin{equation*}
W_{1}\left( y-p,\eta \right) <W_{0}\left( y,\eta \right) <W_{1}\left(
y-p^{\prime },\eta \right) \text{,}
\end{equation*}%
which is a contradiction, since $W_{1}\left( \cdot ,\eta \right) $ is
strictly increasing. In contrast, consider a \textit{continuous} good with
utilities $W\left( x,y-px,\eta \right) $, where $x$ denotes the quantity of
the continuous good, and $W\left( \cdot ,\cdot ,\eta \right) $ is increasing
in both arguments. Now it is possible that $x$ is bought at price $p$ and $%
x^{\prime }$ is bought at price $p^{\prime }$ with $p<p^{\prime }$ and $%
x<x^{\prime }$. That is, we can have%
\begin{equation*}
W\left( x,y-px,\eta \right) <W\left( x^{\prime },y-p^{\prime }x^{\prime
},\eta \right) \text{,}
\end{equation*}%
if $x^{\prime }$ is preferred sufficiently over $x$. The intuitive reason
for this difference between the discrete and the continuous case is that in
the former, the only non-zero option is 1. Indeed, in the continuous case,
it is also not possible that $W\left( x,y-px,\eta \right) <W\left(
x,y-p^{\prime }x,\eta \right) $ for any \textit{common} $x$ if $p<p^{\prime
} $.

Also, note that although Giffen behavior cannot arise in binary choice,
there is no restriction on the \textit{sign of the income effect}. Indeed, (%
\ref{3}) and (\ref{2}) are compatible with both $\frac{\partial }{\partial y}%
\bar{q}\left( p,y\right) \geq 0$ and $\frac{\partial }{\partial y}\bar{q}%
\left( p,y\right) \leq 0$.\medskip

\textbf{D. Parametric and Semiparametric Models}: For a probit/logit
specification of the buying decision, viz.%
\begin{equation}
\bar{q}\left( p,y\right) =F\left( \gamma _{0}+\gamma _{1}p+\gamma
_{2}y\right) =F\left( \gamma _{0}+\left( \gamma _{1}+\gamma _{2}\right)
y-\gamma _{1}\left( y-p\right) \right) \text{,}  \label{14}
\end{equation}%
where $F\left( \cdot \right) $ is a strictly increasing CDF, the shape
restrictions of Theorem 1 amount to requiring $\gamma _{1}\leq 0$ and $%
\gamma _{1}+\gamma _{2}\leq 0$. While the first inequality is intuitive, and
simply says that own price effect is negative, the second condition $\gamma
_{1}+\gamma _{2}\leq 0$ is not a priori obvious, and shows the additional
restriction implied by budget-constrained utility maximization. Now,
applying Theorem 1, we obtain%
\begin{eqnarray*}
&&F\left( \gamma _{0}+\left( \gamma _{1}+\gamma _{2}\right) y-\gamma
_{1}\left( y-p\right) \right) \\
&=&\Pr \left( V\leq F\left( \gamma _{0}+\left( \gamma _{1}+\gamma
_{2}\right) y-\gamma _{1}\left( y-p\right) \right) \right) \\
&=&\Pr \left( \frac{F^{-1}\left( V\right) -\gamma _{0}+\gamma _{1}\left(
y-p\right) }{\gamma _{1}+\gamma _{2}}\geq y\right) \text{,}
\end{eqnarray*}%
where $V\simeq U\left( 0,1\right) $,\footnote{%
We implicitly assume that for fixed $y-p$, the function $q\left(
y,y-p\right) $ varies with $y$ somewhere on $S\left( y-p\right) $, and thus $%
\gamma _{1}+\gamma _{2}\neq 0$.} implying the rationalizing utility functions%
\begin{eqnarray*}
W_{1}\left( y-p,V\right) &=&\frac{F^{-1}\left( V\right) -\gamma _{0}}{\gamma
_{1}+\gamma _{2}}+\underset{\geq 0}{\underbrace{\left( \frac{\gamma _{1}}{%
\gamma _{1}+\gamma _{2}}\right) }}\left( y-p\right) \text{,} \\
W_{0}\left( y,V\right) &=&y\text{.}
\end{eqnarray*}

\begin{remark}
Note that since the restrictions $\gamma _{1}\leq 0$ and $\gamma _{1}+\gamma
_{2}\leq 0$ are linear in parameters, it is computationally straightforward
to maximize a globally concave likelihood, such as probit or logit, subject
to these constraints.
\end{remark}

The above discussion also applies to \textit{semiparametric} binary choice
models (c.f. Manski 1975, Han 1987, Klein and Spady 1993) where one need not
specify the exact functional form of $F\left( \cdot \right) $. For example,
the methods of Cavanagh and Sherman (1998) and Bhattacharya (2008), which
only utilize the strict monotonicity of the CDF $F\left( \cdot \right) $,
can be applied to estimate the binary choice model, subject to our sign
restriction and standard scale-normalization, viz. $\gamma _{1}=-1$ and $%
\gamma _{1}+\gamma _{2}\leq 0$, i.e. using the specification that $\bar{q}%
\left( p,y\right) $ is a strictly increasing function of the linear index $%
-p+\gamma _{2}y$ with $\gamma _{2}\leq 1$.\smallskip

\textbf{E. Random Coefficients}: An alternative parametric specification in
this context is a random coefficient structure, popular in IO applications.
It takes the form%
\begin{eqnarray*}
&&\Pr \left( 1|price=p,income=y\right) \\
&=&\int F\left( \gamma _{1}p+\gamma _{2}y\right) dG\left( \gamma _{1},\gamma
_{2},\theta \right) \\
&=&\int F\left( \left( \gamma _{1}+\gamma _{2}\right) y-\gamma _{1}\left(
y-p\right) \right) dG\left( \gamma _{1},\gamma _{2},\theta \right) \\
&\equiv &H\left( y,y-p,\theta \right) \text{,}
\end{eqnarray*}%
where $\gamma _{1}$ and $\gamma _{2}$ are now random variables with joint
distribution $G\left( \cdot ,\cdot ,\theta \right) $, indexed by an unknown
parameter vector $\theta $, and $F\left( \cdot \right) $ is a specified CDF
(e.g. a probit or logit). Theorem 1 then implies that the distribution $%
G\left( \cdot ,\cdot ,\theta \right) $ must be such that the choice
probability function $H\left( \cdot ,\cdot ,\cdot \right) $ satisfies $\frac{%
\partial }{\partial y}H\left( y,\cdot ,\theta \right) \leq 0$ and $\frac{%
\partial }{\partial \left( y-p\right) }H\left( \cdot ,y-p,\theta \right)
\geq 0$. One way to guarantee this would be to specify the support of $%
\gamma _{1}$ and of $\gamma _{1}+\gamma _{2}$ to lie in $\left( -\infty
,0\right) $. Using Theorem 1, a utility structure that would rationalize
such a model is:%
\begin{equation*}
U_{1}\left( y-p,\eta \right) =h\left( y-p,V,\theta \right) \text{; \ }%
U_{0}\left( y,\eta \right) =y\text{,}
\end{equation*}%
where $V\simeq U\left( 0,1\right) $, and $h\left( y-p,v,\theta \right) $ is $%
\sup \left\{ x:H\left( x,y-p,\theta \right) \geq v\right\} $.\footnote{%
Note that an alternative preference distribution producing the same choice
probabilities is given by $U_{1}\left( y-p,\eta \right) =-\gamma _{1}\left(
y-p\right) $, $U_{0}\left( y,\eta \right) =\gamma _{0}-\left( \gamma
_{1}+\gamma _{2}\right) y$, $\gamma _{0}\perp \left( \gamma _{1},\gamma
_{2}\right) $, $\gamma _{0}\simeq F\left( \cdot \right) $, $\left( \gamma
_{1},\gamma _{2}\right) \simeq G\left( \cdot ,\cdot ,\theta \right) $, $%
\gamma _{1}<0$, $\gamma _{1}+\gamma _{2}\leq 0$ w.p.1. This shows that the
rationalizing preference distribution may not be unique.}

It also follows from the above discussion that not every distribution of
random coefficients $G\left( \cdot ,\cdot ,\theta \right) $ will lead to
rationalizable choice-probability functions. In particular, the commonly
used assumption that $\left( \gamma _{1},\gamma _{2}\right) $ is bivariate
normal (so that the support of $\gamma _{1}$ and of $\gamma _{1}+\gamma _{2}$
do not lie in $\left( -\infty ,0\right) $), can lead to choice probability
functions $H\left( \cdot ,\cdot ,\theta \right) $ that would violate the
shape restrictions of Theorem 1, and thus are not rationalizable.\footnote{%
As a numerical illustration, consider a random coefficient probit model 
\begin{equation*}
\Pr \left( 1|price=p,income=y\right) =\int \Phi \left( \gamma _{1}p+\gamma
_{2}y\right) dF\left( \gamma _{1},\gamma _{2},\theta \right)
\end{equation*}%
where $\gamma _{1}\sim N\left( -1,0.1^{2}\right) $, $\gamma _{2}\sim N\left(
3,0.2^{2}\right) $ and $\gamma _{1}\perp \gamma _{2}$, implying each of the
probabilities of $\gamma _{1}\leq 0$ and $\gamma _{2}\geq 0$ exceeds 0.9999.
Yet it can be verified numerically that e.g.%
\begin{eqnarray*}
&&\frac{\partial }{\partial p}\bar{q}\left( p,y\right) +\frac{\partial }{%
\partial y}\bar{q}\left( p,y\right) |_{p=1,y=1.2} \\
&=&E\left[ \left( \gamma _{1}+\gamma _{2}\right) \times \phi \left( \gamma
_{1}+1.2\times \gamma _{2}\right) \right] \simeq 0.03>0\text{.}
\end{eqnarray*}%
}\medskip

\textbf{F. Observed Covariates}: One can accommodate observed covariates in
our theorem. For example, let $X$ denote a vector of observed covariates,
and let $\bar{q}\left( p,y,x\right) \equiv q\left( y,y-p,x\right) $ denote
the choice probability when $Y=y$, $Y-P=y-p$ and $X=x$. If for each fixed $x$%
, $q\left( \cdot ,\cdot ,x\right) $ satisfies the same properties as (I) A-C
in the statement of Theorem 1, then letting%
\begin{equation*}
q^{-1}\left( u,y-p,x\right) \overset{def}{=}\sup \left\{ z:q\left(
z,y-p,x\right) \geq u\right\} \text{,}
\end{equation*}%
we can rationalize the choice probabilities $\bar{q}\left( p,y,x\right) $ by
setting $W_{1}\left( y-p,V,x\right) \equiv q^{-1}\left( V,y-p,x\right) $ and 
$W_{0}\left( y,V,x\right) \equiv y$, where $V\simeq U\left( 0,1\right) $%
.\medskip

\textbf{G. Endogeneity}: Our results in Theorem 1 are stated in terms of 
\textit{structural} choice probabilities $q\left( \cdot ,\cdot \right) $. If
budget sets are independent of unobserved heterogeneity (conditional on
observed covariates), then these structural choice probabilities are equal
to the observed conditional choice probabilities, i.e.,%
\begin{equation*}
q\left( y,y-p\right) =\Pr \left( 1|Y=y,Y-P=y-p\right) \text{.}
\end{equation*}%
Early results on rationalizability of demand under heterogeneity, including
McFadden and Richter 1990 and Lewbel 2001 worked under such independence. If
the independence condition is violated (even conditional on observed
covariates), then Theorem 1 continues to remain valid as stated, since it
concerns the structural choice probability $q\left( \cdot ,\cdot \right) $,
but consistent estimation of $q\left( \cdot ,\cdot \right) $ will be more
involved. In applications, if endogeneity of budget sets is a potential
concern, then it would be advisable to estimate structural
choice-probabilities using methods for estimating average structural
functions. A specific example is the method of control functions, c.f.
Blundell and Powell 2003, 2004 and Imbens and Newey 2009, which require that 
$\eta \perp \left( P,Y\right) |V$, where $V$ is an estimable
\textquotedblleft control function\textquotedblright\ -- typically a first
stage residual from a regression of endogenous covariates on instruments.
The structural choice probability function can then be recovered (under
regularity conditions) as the integral of the conditional choice probability
given $p,y$ and realizations $v$ of the control variable $V$ over the
marginal distribution of $V$. Hoderlein 2011, Hoderlein and Stoye 2014,
Hausman and Newey 2016, and Kitamura and Stoye 2018 have previously
discussed using control functions to estimate demand nonparametrically.

\section{Empirical Implications}

A practical implication of Theorem 1 is that it can be used to bound
predicted choice probabilities on counterfactual, i.e. previously
unobserved, budget-sets, e.g. those arising from a potential policy
intervention. Such predictions are more reliable when made
nonparametrically, i.e. without arbitrary functional-form/distributional
assumptions on unobservables, and instead based solely on economic
rationality. We now show how to obtain these nonparametric bounds using
Theorem 1.

\textbf{Counterfactual Demand Bounds}: Let $\Omega $ denote the domain of
definition of $\bar{q}\left( \cdot ,\cdot \right) $. Let $A=\left\{ \left(
p^{j},y^{j}\right) ,j=1,...N\right\} \sqsubset \Omega $ denote the set of $%
\left( p,y\right) $ observed in the data, with corresponding choice
probabilities $\left\{ \bar{q}^{j},j=1,...N\right\} =\left\{ q\left(
y^{j},y^{j}-p^{j}\right) \text{, }\left( p^{j},y^{j}\right) \in A\right\} $,
satisfying condition (A) of our Theorem. Suppose we are required to predict
the probability $\bar{q}\left( p^{\prime },y^{\prime }\right) $ of buying at
a counterfactual (i.e. previously unobserved) price $p^{\prime }$ and income 
$y^{\prime }$ with $\left( p^{\prime },y^{\prime }\right) \in \Omega $ $%
\backslash $ $A$. Then Theorem 1 implies the following bounds on this choice
probability:%
\begin{eqnarray}
\bar{L}\left( p^{\prime },y^{\prime }\right) &=&\left\{ 
\begin{array}{l}
\sup_{\left( p,y\right) \in A:\text{ }y\geq y^{\prime },\text{ }y-p\leq
y^{\prime }-p^{\prime }}\bar{q}\left( p,y\right) \text{, if }\left\{ \left(
p,y\right) \in A:\text{ }y\geq y^{\prime },\text{ }y-p\leq y^{\prime
}-p^{\prime }\right\} \neq \phi \\ 
0\text{, if }\left\{ \left( p,y\right) \in A:\text{ }y\geq y^{\prime },\text{
}y-p\leq y^{\prime }-p^{\prime }\right\} =\phi%
\end{array}%
\right.  \label{1} \\
\bar{U}\left( p^{\prime },y^{\prime }\right) &=&\left\{ 
\begin{array}{l}
\inf_{\left( p,y\right) \in A:\text{ }y\leq y^{\prime },\text{ }y-p\geq
y^{\prime }-p^{\prime }}\bar{q}\left( p,y\right) \text{, if }\left\{ \left(
p,y\right) \in A:\text{ }y\leq y^{\prime },\text{ }y-p\geq y^{\prime
}-p^{\prime }\right\} \neq \phi \\ 
1\text{, if }\left\{ \left( p,y\right) \in A:\text{ }y\leq y^{\prime },\text{
}y-p\geq y^{\prime }-p^{\prime }\right\} =\phi%
\end{array}%
\right. \text{.}  \label{9}
\end{eqnarray}%
The above calculation is extremely simple; for example, the lower bound $%
\bar{L}\left( p^{\prime },y^{\prime }\right) $ requires collecting those
observed budget sets $\left( p,y\right) $ in the data that satisfy $y\geq
y^{\prime },$ $y-p\leq y^{\prime }-p^{\prime }$ (a one-line command in
STATA), evaluating choice probabilities on them, and sorting these values.

Note also that for all $\left( p,y\right) \in A$, we have that $\bar{L}%
\left( p,y\right) =\bar{q}\left( p,y\right) =\bar{U}\left( p,y\right) $.

\begin{proposition}
The bounds (\ref{1}) and (\ref{9}) are sharp.
\end{proposition}

\begin{proof}[Proof of Proposition]
Define $W=\left\{ \left( y,y-p\right) :\left( p,y\right) \in A\right\} \cup
\left( y^{\prime },y^{\prime }-p^{\prime }\right) $. Set $\bar{q}\left(
p^{\prime },y^{\prime }\right) \equiv q\left( y^{\prime },y^{\prime
}-p^{\prime }\right) =c$ for any $c$ belonging to the interval defined by
the bounds in (\ref{1}) and (\ref{9}). Then the elements of the set $\left\{
q\left( y,y-p\right) :\left( p,y\right) \in A\cup \left( p^{\prime
},y^{\prime }\right) \right\} $ satisfy the shape restrictions (A) of
Theorem 1 on $W$. In particular, if $\left( p,y\right) \in A$ satisfies $%
y>y^{\prime },$ $y-p=y^{\prime }-p^{\prime }$, then%
\begin{eqnarray*}
\bar{q}\left( p,y\right) &\equiv &q\left( y,y-p\right) \\
&\leq &\sup_{\left( \tilde{p},\tilde{y}\right) \in A:\text{ }\tilde{y}\geq
y^{\prime },\text{ }\tilde{y}-\tilde{p}\leq y^{\prime }-p^{\prime }}q\left( 
\tilde{y},\tilde{y}-\tilde{p}\right) \text{, since }q\left( \cdot ,\cdot
\right) \text{ satisfies cond (A) of Thm 1 on }A \\
&\equiv &\sup_{\left( \tilde{p},\tilde{y}\right) \in A:\text{ }\tilde{y}\geq
y^{\prime },\text{ }\tilde{y}-\tilde{p}\leq y^{\prime }-p^{\prime }}\bar{q}%
\left( \tilde{p},\tilde{y}\right) \\
&\leq &c=\bar{q}\left( p^{\prime },y^{\prime }\right) \text{;}
\end{eqnarray*}%
on the other hand, if $\left( p,y\right) \in A$ satisfies $y=y^{\prime },$ $%
y-p>y^{\prime }-p^{\prime }$, then%
\begin{eqnarray*}
\bar{q}\left( p,y\right) &\equiv &q\left( y,y-p\right) \\
&\geq &\inf_{\left( \tilde{p},\tilde{y}\right) \in A:\text{ }\tilde{y}\leq
y^{\prime },\text{ }\tilde{y}-\tilde{p}\geq y^{\prime }-p^{\prime }}q\left( 
\tilde{y},\tilde{y}-\tilde{p}\right) \text{, since }q\left( \cdot ,\cdot
\right) \text{ satisfies cond (A) of Thm 1 on }A \\
&\equiv &\inf_{\left( \tilde{p},\tilde{y}\right) \in A:\text{ }\tilde{y}\leq
y^{\prime },\text{ }\tilde{y}-\tilde{p}\geq y^{\prime }-p^{\prime }}\bar{q}%
\left( \tilde{p},\tilde{y}\right) \geq c=\bar{q}\left( p^{\prime },y^{\prime
}\right) \text{.}
\end{eqnarray*}%
Next, note that conditions (B) and (C) of our theorem have no empirical
content vis-a-vis the countably finite set of values $\left\{ \bar{q}^{j}%
\text{, }j=1,...,N\right\} \cup \left\{ c\right\} $, in that there are no
set of values $\left\{ \bar{q}^{j}\text{, }j=1,...,N\right\} \cup \left\{
c\right\} $ which can imply a violation of conditions (B) and (C).
Therefore, the choice probabilities $\left\{ \bar{q}^{j}\text{, }%
j=1,...,N\right\} \cup \left\{ c\right\} $ corresponding to $A\cup \left(
p^{\prime },y^{\prime }\right) $ are compatible with a choice probability
function $q\left( \cdot ,\cdot \right) $ on a domain $G$ containing $W\cup
\left( y^{\prime },y^{\prime }-p^{\prime }\right) $ and satisfying
conditions (A)-(C) of Theorem 1 (for an explicit construction of such a\
function, see discussion on discrete support of $\left( P,Y\right) $ in the
paragraph preceding Theorem 1 above). Therefore, applying Theorem 1, we
conclude that there exist utility functions $W_{1}\left( a_{1},V\right) $
and $W_{0}\left( a_{0},V\right) =a_{0}$ with $V\simeq U\left( 0,1\right) $
that satisfy the restrictions (A')-(C') of Theorem 1, and $\Pr \left[
W_{1}\left( a_{1},V\right) \geq a_{0}\right] =q\left( a_{0},a_{1}\right) $
for all $\left( a_{0},a_{1}\right) \in G$; in particular,%
\begin{eqnarray*}
\Pr \left[ W_{1}\left( y^{j}-p^{j},V\right) \geq y^{j}\right] &=&q^{j}\text{%
, }j=1,...,N\text{,} \\
\text{and }\Pr \left[ W_{1}\left( y^{\prime }-p^{\prime },V\right) \geq
y^{\prime }\right] &=&c\text{.}
\end{eqnarray*}
\end{proof}

\textbf{Welfare bounds}: Given bounds on choice probabilities, one can
obtain lower and upper bounds on economically interesting functionals
thereof, such as average welfare. For example, the average compensating
variation -- i.e. utility preserving income compensation -- corresponding to
a price increase from $p_{0}$ to $p_{1}$ at income $y$ is given by $%
\int_{p_{0}}^{p_{1}}\bar{q}\left( p,y+p-p_{0}\right) dp$ (c.f. Bhattacharya
2015). This requires prediction of demand on a continuum of budget sets,
viz. $\left\{ \bar{q}\left( p,y+p-p_{0}\right) :p\in \left[ p_{0},p_{1}%
\right] \right\} $. Now, it follows from our discussion immediately above,
and by Theorem 1, that pointwise sharp bounds on $\bar{q}\left(
p,y+p-p_{0}\right) $ are given by%
\begin{eqnarray}
&&\bar{L}\left( p,y+p-p_{0}\right)  \notag \\
&\equiv &\left\{ 
\begin{array}{l}
\sup_{\left( \tilde{p},\tilde{y}\right) \in A\text{, }\tilde{y}-\tilde{p}%
\leq y-p_{0}\text{, }\tilde{y}\geq y+p-p_{0}}\bar{q}\left( \tilde{p},\tilde{y%
}\right) \text{, if }\left\{ \left( \tilde{p},\tilde{y}\right) \in A\text{, }%
\tilde{y}-\tilde{p}\leq y-p_{0}\text{, }\tilde{y}\geq y+p-p_{0}\right\} \neq
\phi \\ 
0\text{, if }\left\{ \left( \tilde{p},\tilde{y}\right) \in A\text{, }\tilde{y%
}-\tilde{p}\leq y-p_{0}\text{, }\tilde{y}\geq y+p-p_{0}\right\} =\phi%
\end{array}%
\right.  \notag \\
&\leq &\bar{q}\left( p,y+p-p_{0}\right)  \notag \\
&\leq &\left\{ 
\begin{array}{l}
\inf_{\left( \tilde{p},\tilde{y}\right) \in A\text{, }\tilde{y}-\tilde{p}%
\geq y-p_{0}\text{, }\tilde{y}\leq y+p-p_{0}}\bar{q}\left( \tilde{p},\tilde{y%
}\right) \text{, if }\left\{ \left( \tilde{p},\tilde{y}\right) \in A\text{, }%
\tilde{y}-\tilde{p}\geq y-p_{0}\text{, }\tilde{y}\leq y+p-p_{0}\right\} \neq
\phi \\ 
1\text{, if }\left\{ \left( \tilde{p},\tilde{y}\right) \in A\text{, }\tilde{y%
}-\tilde{p}\geq y-p_{0}\text{, }\tilde{y}\leq y+p-p_{0}\right\} =\phi%
\end{array}%
\right.  \notag \\
&\equiv &\bar{M}\left( p,y+p-p_{0}\right) \text{.}  \label{5}
\end{eqnarray}%
This implies that average CV at $y$ is bounded below by $\int_{p_{0}}^{p_{1}}%
\bar{L}\left( p,y+p-p_{0}\right) dp$, and above by $\int_{p_{0}}^{p_{1}}\bar{%
M}\left( p,y+p-p_{0}\right) dp$.

As for sharpness, let $L\left( y,y-p\right) =\bar{L}\left( p,y\right) $ be
defined analogous to $q\left( y,y-p\right) =\bar{q}\left( p,y\right) $
above. Then the lower bound on average CV becomes $\int_{p_{0}}^{p_{1}}L%
\left( y+p-p_{0},y-p_{0}\right) $. Now, by definition,%
\begin{eqnarray*}
&&L\left( a_{0},a_{1}\right) \\
&=&\left\{ 
\begin{array}{l}
\sup \left\{ \bar{q}\left( \tilde{p},\tilde{y}\right) :\left( \tilde{p},%
\tilde{y}\right) \in A\text{, }\tilde{y}-\tilde{p}\leq a_{1}\text{, }\tilde{y%
}\geq a_{0}\right\} \text{, if }\left\{ \left( \tilde{p},\tilde{y}\right)
\in A\text{, }\tilde{y}-\tilde{p}\leq a_{1}\text{, }\tilde{y}\geq
a_{0}\right\} \neq \phi \\ 
0\text{, if }\left\{ \left( \tilde{p},\tilde{y}\right) \in A\text{, }\tilde{y%
}-\tilde{p}\leq a_{1}\text{, }\tilde{y}\geq a_{0}\right\} =\phi%
\end{array}%
\right.
\end{eqnarray*}%
is non-increasing in $a_{0}$ and non-decreasing in $a_{1}$, and $L\left(
y,y-p\right) =q\left( y,y-p\right) $ when $\left( p,y\right) \in A$.
Furthermore, for fixed value of $\left( y-p_{0}\right) $, as $p$ varies over
the interval $\left[ p_{0},p_{1}\right] $, the function $L\left(
y+p-p_{0},y-p_{0}\right) $ can assume at most finitely many values (viz. $%
q\left( y^{m},y^{m}-p^{m}\right) $, $m=1,...,N$), and therefore, must
necessarily be piecewise flat in $p$, with at most countably finite number
of discontinuity points. Therefore, one can construct a function $Q\left(
\cdot ,\cdot \right) $ (see footnote below for an illustration) that (1) is
continuous in the first argument, (2) equals $L\left( \cdot ,\cdot \right) $
(and therefore $q\left( \cdot ,\cdot \right) $) on $A$, (3) equals $L\left(
\cdot ,\cdot \right) $ everywhere else on the domain except in arbitrarily
small (semi-closed) intervals around the points of discontinuity of $L\left(
\cdot ,\cdot \right) $, and (4) satisfies the same shape restrictions as $%
L\left( \cdot ,\cdot \right) $ ; also, (5) $Q\left( \cdot ,\cdot \right) $
can be trivially made to satisfy the limit conditions (C) of Theorem 1 by
defining the limit points $y_{L}\left( \cdot \right) $, $y_{H}\left( \cdot
\right) $ lower than the lowest and larger than the highest values
respectively attained by $y$ in $A$ corresponding to any fixed value of $y-p$%
. Using (1), (4) and (5) and applying Theorem 1, we can rationalize $Q\left(
\cdot ,\cdot \right) $ -- which equals $q\left( \cdot ,\cdot \right) $ at
all the observed data points, i.e. corresponding to $\left( p,y\right) \in A$
-- via a pair of utility functions and a uniformly distributed unobserved
heterogeneity, and at the same time, $\int_{p_{0}}^{p_{1}}Q\left(
y+p-p_{0},y-p_{0}\right) dp$, is arbitrarily close to $\int_{p_{0}}^{p_{1}}L%
\left( y+p-p_{0},y-p_{0}\right) =\int_{p_{0}}^{p_{1}}\bar{L}\left(
p,y+p-p_{0}\right) dp$, since they differ only on at most finitely many
intervals of arbitrarily small length. Therefore, $\int_{p_{0}}^{p_{1}}\bar{L%
}\left( p,y+p-p_{0}\right) dp$ is a sharp lower bound for average CV $%
\int_{p_{0}}^{p_{1}}q\left( y+p-p_{0},y-p_{0}\right) \equiv
\int_{p_{0}}^{p_{1}}\bar{q}\left( p,y+p-p_{0}\right) dp$.\footnote{%
As a simple illustration, consider a fixed $a_{1}=y-p_{0}\in \Omega _{1}$,
and suppose the point $\left( k,a_{1}\right) \in A$, and $l<k<u$ for some
real numbers $l,u$ belonging to the interval $\left[ y,y+p_{1}-p_{0}\right] $
where the first argument of $L\left( y+p-p_{0},y-p_{0}\right) $ takes its
values as $p$ varies over $\left[ p_{0},p_{1}\right] $. Now suppose the
lower bound function $L\left( \cdot .\cdot \right) $ satisfies 
\begin{equation*}
L\left( a_{0},a_{1}\right) =\left\{ 
\begin{array}{l}
q\left( k,a_{1}\right) \text{ if }l\leq a_{0}\leq k \\ 
L\left( k^{+},a_{1}\right) \text{ if }k<a_{0}\leq u%
\end{array}%
\right.
\end{equation*}%
with $L\left( k^{+},a_{1}\right) <q\left( k,a_{1}\right) $. That is, $%
L\left( \cdot ,\cdot \right) $ equals $q\left( \cdot ,\cdot \right) $ at the
point $\left( k,a_{1}\right) $ in $A$, is non-increasing in the first
argument and is (right) discontinuous at $k$ with $L\left(
k^{+},a_{1}\right) <L\left( k,a_{1}\right) $. Choose $\delta \in \left(
0,u-k\right) $ and define the function $Q\left( \cdot ,a_{1}\right) $ as%
\begin{equation*}
Q\left( a_{0},a_{1}\right) =\left\{ 
\begin{array}{l}
L\left( k,a_{1}\right) \text{, if }l\leq a_{0}\leq k \\ 
L\left( k,a_{1}\right) \times \left[ 1-\frac{a_{0}-k}{\delta }\right]
+L\left( k^{+},a_{1}\right) \frac{a_{0}-k}{\delta }\text{ if }k<a_{0}\leq
k+\delta \\ 
L\left( k^{+},a_{1}\right) \text{, if }k+\delta <a_{0}\leq u%
\end{array}%
\right.
\end{equation*}%
Then (1) $Q\left( \cdot ,a_{1}\right) $ is continuous in the first argument,
since $Q\left( a_{0},a_{1}\right) \nearrow L\left( k,a_{1}\right) $ as $%
a_{0}\searrow k$, and $\searrow L\left( k^{+},a_{1}\right) $ as $%
a_{0}\nearrow \left( k+\delta \right) $, (2) at the point $\left(
k,a_{1}\right) \in A$, $Q\left( k,a_{1}\right) =q\left( k,a_{1}\right)
=L\left( k,a_{1}\right) $, (3) $Q\left( \cdot ,a_{1}\right) $ equals $%
L\left( \cdot ,a_{1}\right) $ except on the semi-open interval $(k,k+\delta
] $ of length $\delta $, (4) $Q\left( \cdot ,a_{1}\right) $ is
non-increasing, and $Q\left( a_{0},\cdot \right) $. is non-decreasing since $%
L\left( \cdot ,a_{1}\right) $ is non-increasing, and $L\left( a_{0},\cdot
\right) $. is non-decreasing. Finally, $\int_{l}^{u}Q\left(
a_{0},a_{1}\right) da_{0}-\int_{l}^{u}L\left( a_{0},a_{1}\right) da_{0}$
equals the area of the triangle with base $\delta $ and height $L\left(
k,a_{1}\right) -L\left( k^{+},a_{1}\right) $ thus equalling $\frac{L\left(
k,a_{1}\right) -L\left( k^{+},a_{1}\right) }{2}\delta $ which can be made
arbitrarily close to 0 by choosing $\delta $ arbitrarily close to 0.}

A symmetric line of argument implies that $\int_{p_{0}}^{p_{1}}\bar{M}\left(
p,y+p-p_{0}\right) dp$ is the sharp upper bound.

\section{Connection with Revealed Stochastic Preference}

The welfare calculation above requires prediction of demand on a continuum
of budget sets indexed by $p\in \left[ p_{0},p_{1}\right] $, which is
operationally difficult -- if not practically impossible -- to implement,
using the finite-dimensional matrix equation based SRP approach. But in
simple cases where there are a small, countably finite number of budget
sets, and it is easy to verify the SRP conditions, a natural question is
whether our shape restrictions (A) of Theorem 1 are compatible with the SRP
based criterion for rationalizability; condition (B) and (C) of Theorem 1
are of course irrelevant in such cases. Below, we show that our shape
restrictions (A) are in fact \textit{necessary} for the SRP criterion to be
satisfied.

\begin{proposition}
The shape restrictions (A) in Theorem 1 are necessary for McFadden Richter's
SRP conditions to hold.
\end{proposition}

\begin{proof}
Consider two price and income combinations $\left( p^{1},y\right) $ and $%
\left( p^{2},y\right) $. Suppose WLOG that $p^{1}\,<p^{2}$, i.e., $%
y-p^{1}>y-p^{2}$. Let $q\left( y,y-p^{1}\right) $, $q\left( y,y-p^{2}\right) 
$ denote choice probabilities of alternative 1 on the two budgets,
respectively. Assume, if possible, that out shape restriction A(ii) is
violated, so that $q\left( y,y-p^{1}\right) <q\left( y,y-p^{2}\right) $. We
will show that this implies violation of McFadden-Richter's SRP condition.
Toward that end, consider three bundles $\left( 0,y\right) ,\left(
1,y-p^{1}\right) $ and $\left( 1,y-p^{2}\right) $. Under nonsatiation in
numeraire, there are 3 possible preference profiles in the population, given
by (i) $\left( 0,y\right) \succ \left( 1,y-p^{1}\right) \succ \left(
1,y-p^{2}\right) $, (ii) $\left( 1,y-p^{1}\right) \succ \left( 0,y\right)
\succ \left( 1,y-p^{2}\right) $ and (iii) $\left( 1,y-p^{1}\right) \succ
\left( 1,y-p^{2}\right) \succ \left( 0,y\right) $; assume the population
proportions of these three profiles are $\left( \pi _{1},\pi _{2},\pi
_{3}\right) $, respectively. Then McFadden-Richter's SRP condition is that
the matrix equation%
\begin{eqnarray}
\left[ 
\begin{array}{ccc}
0 & 1 & 1 \\ 
0 & 0 & 1%
\end{array}%
\right] \left[ 
\begin{array}{c}
\pi _{1} \\ 
\pi _{2} \\ 
\pi _{3}%
\end{array}%
\right] &=&\left[ 
\begin{array}{c}
q\left( y,y-p^{1}\right) \\ 
q\left( y,y-p^{2}\right)%
\end{array}%
\right] \text{, i.e.}  \notag \\
\pi _{2}+\pi _{3} &=&q\left( y,y-p^{1}\right) \text{, }\pi _{3}=q\left(
y,y-p^{2}\right) \text{,}  \label{6}
\end{eqnarray}%
has a solution $\left( \pi _{1},\pi _{2},\pi _{3}\right) $ in the unit
positive simplex. But if our hypothesis holds, i.e. $q\left(
y,y-p^{1}\right) <q\left( y,y-p^{2}\right) $, then (\ref{6}) implies $\pi
_{2}+\pi _{3}<\pi _{3}$ i.e. $\pi _{2}<0$, a violation.

Next, consider the two price and income combinations $\left(
p^{1},y^{1}\right) $ and $\left( p^{2},y^{2}\right) $ with $y^{1}<y^{2}$ and 
$y^{1}-p^{1}=y^{2}-p^{2}\equiv a_{1}$, say. Let $q\left( y^{1},a_{1}\right) $%
, $q\left( y^{2},a_{1}\right) $ denote choice probabilities of alternative 1
on the two budgets, respectively. Now suppose our shape restriction A(i) is
violated, so that $q\left( y^{1},a_{1}\right) <q\left( y^{2},a_{1}\right) $.
Consider the three bundles $\left( 0,y^{1}\right) $, $\left( 0,y^{2}\right) $
and $\left( 1,a_{1}\right) $. Under nonsatiation, there are 3 possible
preference profiles in the population, given by (i) $\left( 0,y^{2}\right)
\succ \left( 0,y^{1}\right) \succ \left( 1,a_{1}\right) $, (ii) $\left(
0,y^{2}\right) \succ \left( 1,a_{1}\right) \succ \left( 0,y^{1}\right) $ and
(iii) $\left( 1,a_{1}\right) \succ \left( 0,y^{2}\right) \succ \left(
0,y^{1}\right) $; assume the population proportions of these three profiles
are $\left( \pi _{1},\pi _{2},\pi _{3}\right) $, respectively. Then SRP
requires a solution $\left( \pi _{1},\pi _{2},\pi _{3}\right) $ in the unit
positive simplex to%
\begin{eqnarray}
\left[ 
\begin{array}{ccc}
0 & 1 & 1 \\ 
0 & 0 & 1%
\end{array}%
\right] \left[ 
\begin{array}{c}
\pi _{1} \\ 
\pi _{2} \\ 
\pi _{3}%
\end{array}%
\right] &=&\left[ 
\begin{array}{c}
q\left( y^{1},a_{1}\right) \\ 
q\left( y^{2},a_{1}\right)%
\end{array}%
\right] \text{, i.e.}  \notag \\
\pi _{2}+\pi _{3} &=&q\left( y^{1},a_{1}\right) \text{, }\pi _{3}=q\left(
y^{2},a_{1}\right) \text{.}  \label{8}
\end{eqnarray}%
But $q\left( y^{1},a_{1}\right) <q\left( y^{2},a_{1}\right) $ implies that $%
\pi _{2}+\pi _{3}<\pi _{3}$ implying $\pi _{2}<0$, which is a violation of $%
\left( \pi _{1},\pi _{2},\pi _{3}\right) $ lying in the unit positive
simplex.
\end{proof}

With more budget sets, the corresponding higher dimensional matrix equations
analogous to (\ref{6}) and (\ref{8}) quickly become operationally
impractical and cumbersome, as is well-known in the literature (see
introduction). In contrast, our shape-restrictions, by being global
conditions on the $q\left( \cdot ,\cdot \right) $ functions, remain
invariant to which and how many budget sets are considered. Furthermore, we
already know via Theorem 1 above, that these shape restrictions are also 
\textit{sufficient} for rationalizability for \textit{any} collection --
finite or infinite -- of budget sets.\footnote{%
It does not seem possible to show directly, i.e. \textit{without using
Theorem 1}, that our shape restrictions are also \textit{sufficient} for
existence of admissible solutions to the analog of (\ref{6}) and (\ref{8})
corresponding to \textit{every} arbitrary collection of budget sets. But
given theorem 1, this exercise is probably of limited interest.}

\begin{center}
\textbf{Appendix\medskip }

\textbf{1. Proof of Theorem 1}
\end{center}

\begin{proof}[Proof]
That (II) implies (I) is straightforward. In particular, letting $%
W_{0}^{-1}\left( \cdot ,\eta \right) $ denote the inverse of $W_{0}\left(
\cdot ,\eta \right) $, we have that 
\begin{equation*}
q\left( y,y-p\right) =\int 1\left\{ y\leq W_{0}^{-1}\left( W_{1}\left(
y-p,\eta \right) ,\eta \right) \right\} dG\left( \eta \right)
\end{equation*}%
whence (B') implies (B), (C') implies (C), and (A') implies (A).

We now show that (I) implies (II).

Note that (C) implies that for any $v\in \left[ 0,1\right] $ and $a_{1}\in
\Omega _{1}$, the set $\left\{ a_{0}\in \left[ y_{L}\left( a_{1}\right)
,y_{H}\left( a_{1}\right) \right] :q\left( a_{0},a_{1}\right) \geq v\right\} 
$ is non-empty; for any fixed $a_{1}\in \Omega _{1}$ and for $v\in \left[ 0,1%
\right] $, define%
\begin{equation}
q^{-1}\left( v,a_{1}\right) \overset{def}{=}\sup \left\{ a_{0}\in \left[
y_{L}\left( a_{1}\right) ,y_{H}\left( a_{1}\right) \right] :q\left(
a_{0},a_{1}\right) \geq v\right\} \text{,}  \label{4}
\end{equation}%
which takes values in $\left[ y_{L}\left( a_{1}\right) ,y_{H}\left(
a_{1}\right) \right] $.\footnote{%
Here we are implicitly assuming that $\Omega _{0}\left( a_{1}\right) $
equals (or contains) $[y_{L}\left( a_{1}\right) ,y_{H}\left( a_{1}\right) ]$%
. If however the support of price and income are discrete, then $\Omega
_{0}\left( a_{1}\right) $ can be a strict subset of $[y_{L}\left(
a_{1}\right) ,y_{H}\left( a_{1}\right) ]$. Then $q\left( \cdot ,\cdot
\right) $ is not defined at the points `in between' the points of support,
and therefore, $q^{-1}\left( \cdot ,a_{1}\right) $ in (\ref{4}) is not
well-defined. To cover this case, one can extend $q\left( \cdot ,\cdot
\right) $ to a continuous function $q^{c}\left( \cdot ,\cdot \right) $
defined on a rectangle $\Omega ^{c}$ containing $\Omega $ such that (i) $%
q^{c}\left( \cdot ,\cdot \right) $ equals $q\left( \cdot ,\cdot \right) $ on 
$\Omega $, (ii) $q^{c}\left( \cdot ,\cdot \right) $ satisfies the same shape
restrictions on $\Omega ^{c}$ that are satisfied by $q\left( \cdot ,\cdot
\right) $ on $\Omega $, and (iii) $q^{c}\left( \cdot ,\cdot \right) $
satisfies the limit conditions C of Theorem 1. In the online appendix, we
provide an explicit construction of such a function. The proof of Theorem 1
then holds with $\Omega $, $\Omega _{0}\left( \cdot \right) $ and $q\left(
\cdot ,\cdot \right) $ equalling their corresponding extensions in the case
where $\left( P,Y\right) $ have discrete support.} Also, by condition (A), $%
q^{-1}\left( v,\cdot \right) $ must be non-decreasing.

Now, consider a random variable $V\simeq Uniform\left( 0,1\right) $. Define $%
W_{0}\left( a_{0},V\right) \overset{defn}{=}a_{0}$ and $W_{1}\left(
a_{1},V\right) \overset{defn}{=}q^{-1}\left( V,a_{1}\right) $. We will now
show that $W_{0}\left( \cdot ,V\right) $ and $W_{1}\left( \cdot ,V\right) $
will rationalize the choice-probabilities $q\left( \cdot ,\cdot \right) $,
and satisfy properties (A')-(C') of our theorem.

To do so, first note that for any fixed $a_{1}\in \Omega _{1}$, the function 
$1-q\left( \cdot ,a_{1}\right) $ is a continuous CDF by conditions A(i), B
and C of the theorem, and $q^{-1}\left( v,a_{1}\right) $ is, by definition,
the corresponding $\left( 1-v\right) $th quantile. Standard properties of
quantiles, c.f. Pfeiffer 1990, Sec 11a, Page 266-7, then imply the following
three results (for completeness, we state and prove these results formally
as a Claim below this proof):

\textbf{Result (i)}: for any $a_{1}\in \Omega _{1}$ and $v\in \left[ 0,1%
\right] $, we must have that $q\left( q^{-1}\left( v,a_{1}\right)
,a_{1}\right) =v$ (Pfeiffer 1990, page 267, property 6);

\textbf{Result (ii)}: for any $a_{1}\in \Omega _{1}$, $a_{0}\in \left[
y_{L}\left( a_{1}\right) ,y_{H}\left( a_{1}\right) \right] $ and $v\in \left[
0,1\right] $, we have $q\left( a_{0},a_{1}\right) \geq v\Leftrightarrow $ $%
a_{0}\leq q^{-1}\left( v,a_{1}\right) $ (Pfeiffer 1990 page 266 property 1);

\textbf{Result (iii)}: for any $a_{1}\in \Omega _{1}$, the function $%
q^{-1}\left( \cdot ,a_{1}\right) $ is one-to-one on $\left[ 0,1\right] $
(Consequence of \textbf{Result (i)}).\smallskip

Now, for $V\simeq Uniform\left( 0,1\right) $, it follows from \textbf{Result
(ii)} that%
\begin{equation}
\Pr \left( q^{-1}\left( V,a_{1}\right) \geq a_{0}\right) =\Pr \left( V\leq
q\left( a_{0},a_{1}\right) \right) =q\left( a_{0},a_{1}\right) \text{.}
\label{X}
\end{equation}%
Therefore, the utility functions $W_{0}\left( a_{0},V\right) \equiv a_{0}$
and $W_{1}\left( a_{1},V\right) \equiv q^{-1}\left( V,a_{1}\right) $ with
heterogeneity $V\simeq Uniform\left( 0,1\right) $ rationalize the choice
probabilities $q\left( \cdot ,\cdot \right) $, and satisfy all the
properties specified in panel (II) of Theorem 1. In particular, $W_{1}\left(
a_{1},\eta \right) $ is non-decreasing in $a_{1}$ (see right after eqn. (\ref%
{4})), so (A'ii) holds; $W_{0}\left( a_{0},\eta \right) =a_{0}$ trivially
satisfies (A'i). Next, for $v,v^{\prime }\in \left[ 0,1\right] $ with $v\neq
v^{\prime }$, we cannot have that $q^{-1}\left( v,a_{1}\right) =q^{-1}\left(
v^{\prime },a_{1}\right) $ by \textbf{Result (iii)}; therefore,%
\begin{equation}
\Pr \left[ q^{-1}\left( V,a_{1}\right) =a_{0}\right] =0\text{ for all }a_{0}%
\text{,}  \label{XX}
\end{equation}%
which implies property (B'). Finally,%
\begin{eqnarray*}
&&\lim_{y\searrow y_{L}\left( a_{1}\right) ,y-p=a_{1}}\Pr \left[
q^{-1}\left( V,y-p\right) \geq y\right] \\
&&\overset{\text{by (\ref{X})}}{=}\lim_{y\searrow y_{L}\left( a_{1}\right)
,y-p=a_{1}}\Pr \left[ q\left( y,y-p\right) \geq V\right] \\
&=&\lim_{y\searrow y_{L}\left( a_{1}\right) ,y-p=a_{1}}q\left( y,y-p\right) 
\text{, since }V\simeq U\left( 0,1\right) \\
&&\overset{\text{by Condition (C)}}{=}1\text{.}
\end{eqnarray*}%
By an analogous argument, $\lim_{y\nearrow y_{H}\left( a_{1}\right)
,y-p=a_{1}}\Pr \left[ q^{-1}\left( V,y-p\right) \geq y\right] =0$, thus
satisfying (C').\smallskip
\end{proof}

\begin{center}
\textbf{2. Proof of Results (i), (ii) and (iii) in Theorem 1}
\end{center}

\textbf{Claim}: \textit{Suppose }$q\left( \cdot ,\cdot \right) :\Omega
\rightarrow \left[ 0,1\right] $\textit{\ satisfies conditions (A), (B), (C)
of Theorem 1, and }$q^{-1}\left( \cdot ,\cdot \right) $\textit{\ is as
defined in (\ref{4}). Then (i) for any }$a_{1}\in \Omega _{1}$\textit{\ and }%
$v\in \left[ 0,1\right] $\textit{, we must have that }$q\left( q^{-1}\left(
v,a_{1}\right) ,a_{1}\right) =v$\textit{; (ii) for any }$v\in \lbrack 0,1]$%
\textit{, and any }$\left( a_{0},a_{1}\right) \in \Omega $\textit{, we have
that }$q\left( a_{0},a_{1}\right) \geq v\Longleftrightarrow a_{0}\leq
q^{-1}\left( v,a_{1}\right) $\textit{; (iii) for any }$a_{1}\in \Omega _{1}$%
\textit{, the function }$q^{-1}\left( \cdot ,a_{1}\right) $\textit{\ is
one-to-one on }$\left[ 0,1\right] $\textit{.}\smallskip

\begin{proof}[Proof]
\textbf{Claim (i): }Pick $a_{1}\in \Omega _{1}$. For $v=0$, we cannot have
that $q\left( q^{-1}\left( v,a_{1}\right) ,a_{1}\right) <v$, since $q\left(
\cdot ,\cdot \right) $ takes values in $[0,1]$. So let $v\in \mathcal{(}0,1]$%
, and suppose if possible that $q\left( q^{-1}\left( v,a_{1}\right)
,a_{1}\right) <v$. Note that $q^{-1}\left( v,a_{1}\right) >y_{L}\left(
a_{1}\right) $ because if $q^{-1}\left( v,a_{1}\right) =y_{L}\left(
a_{1}\right) $, then $q\left( q^{-1}\left( v,a_{1}\right) \right) =q\left(
y_{L}\left( a_{1}\right) ,a_{1}\right) =1\geq v$. Therefore, $q\left(
q^{-1}\left( v,a_{1}\right) ,a_{1}\right) <v$ implies by the continuity
condition (B) that there must exist $\varepsilon >0$\ such that $q\left(
x,a_{1}\right) <v$\ for all $x\in \left[ q^{-1}\left( v,a_{1}\right)
-\varepsilon ,q^{-1}\left( v,a_{1}\right) \right] $. But by condition (A)
and the definition of $q^{-1}\left( \cdot ,a_{1}\right) $\ as the supremum
in (\ref{4}), we must have that $q\left( x,a_{1}\right) \geq v$\ for all $%
x<q^{-1}\left( v,a_{1}\right) $, and in particular for $x\in \left[
q^{-1}\left( v,a_{1}\right) -\varepsilon ,q^{-1}\left( v,a_{1}\right) \right]
$, which contradicts $q\left( x,a_{1}\right) <v$.

Next, for $v=1$, we cannot have that $q\left( q^{-1}\left( v,a_{1}\right)
,a_{1}\right) >v$, since $q\left( \cdot ,\cdot \right) $ takes values in $%
[0,1]$. So let $v\in \lbrack 0,1)$ and suppose $q\left( q^{-1}\left(
v,a_{1}\right) ,a_{1}\right) >v$. Condition (B) and (C) imply via the
intermediate value theorem that $\ni x\in \Omega _{0}\left( a_{1}\right) $,
such that $q\left( x,a_{1}\right) =v$. But by hypothesis, $q\left(
q^{-1}\left( v,a_{1}\right) ,a_{1}\right) >v=q\left( x,a_{1}\right) $, so
(A) implies that $x>q^{-1}\left( v,a_{1}\right) $, which, together with $%
q\left( x,a_{1}\right) =v$, contradicts $q^{-1}\left( v,a_{1}\right) $ being
the supremum in (\ref{4}). Therefore, $q\left( q^{-1}\left( v,a_{1}\right)
,a_{1}\right) =v$ for all $v\in \left[ 0,1\right] $, and Claim (i) is
proved.\smallskip

\textbf{Claim (ii):} To prove claim (ii), note that for any $v\in \lbrack
0,1]$, and any $\left( a_{0},a_{1}\right) \in \Omega $,%
\begin{equation}
a_{0}\leq q^{-1}\left( v,a_{1}\right) \overset{\text{by (A)}}{%
\Longrightarrow }q\left( a_{0},a_{1}\right) \geq \underset{=v\text{, by
Result (i)}}{\underbrace{q\left( q^{-1}\left( v,a_{1}\right) ,a_{1}\right) }}%
\Longrightarrow q\left( a_{0},a_{1}\right) \geq v\text{.}  \label{a'}
\end{equation}%
Also, by definition of $q^{-1}\left( \cdot ,a_{1}\right) $ as the supremum
in (\ref{4}), we have by (A) that%
\begin{equation}
q\left( a_{0},a_{1}\right) \geq v\Longrightarrow a_{0}\leq q^{-1}\left(
v,a_{1}\right) \text{.}  \label{b'}
\end{equation}%
Therefore, from (\ref{a'}) and (\ref{b'}), we have that $q\left(
a_{0},a_{1}\right) \geq v\Longleftrightarrow a_{0}\leq q^{-1}\left(
v,a_{1}\right) $, which proves claim (ii).\smallskip

\textbf{Claim (iii): }To prove claim (iii), note that for $v,v^{\prime }\in %
\left[ 0,1\right] $ with $v\neq v^{\prime }$, we cannot have that $%
q^{-1}\left( v,a_{1}\right) =q^{-1}\left( v^{\prime },a_{1}\right) $;
otherwise,%
\begin{equation*}
v\overset{\text{by Claim (i)}}{=}q\left( q^{-1}\left( v,a_{1}\right)
,a_{1}\right) \overset{\text{by }q^{-1}\left( v,a_{1}\right) =q^{-1}\left(
v^{\prime },a_{1}\right) }{=}q\left( q^{-1}\left( v^{\prime },a_{1}\right)
,a_{1}\right) \overset{\text{by Claim (i)}}{=}v^{\prime },
\end{equation*}%
contradicting $v\neq v^{\prime }$.\pagebreak
\end{proof}

\begin{center}
\textbf{References}
\end{center}

\begin{enumerate}
\item Anderson, S.P., De Palma, A. and Thisse, J.F. 1992. Discrete choice
theory of product differentiation. MIT press.

\item Bhattacharya, D. 2015. Nonparametric welfare analysis for discrete
choice. Econometrica, 83(2), pp.617-649.

\item Bhattacharya, D. 2018. Empirical welfare analysis for discrete choice:
Some general results. Quantitative Economics, 9(2), pp.571-615.

\item Bhattacharya, D. 2008. A Permutation-Based Estimator for Monotone
Index Models. Econometric Theory 24(3), pp.795-807.

\item Blundell, R., and James L. Powell (2003): Endogeneity in nonparametric
and semiparametric regression models. Econometric society monographs 36,
312-357.

\item Blundell, R.W. and Powell, J.L. (2004): Endogeneity in semiparametric
binary response models. The Review of Economic Studies, 71(3), 655-679.

\item Cavanagh, C. and Sherman, R.P. (1998): Rank estimators for monotonic
index models. Journal of Econometrics, 84(2), 351-382.

\item Chetverikov, D. (2012): Testing regression monotonicity in econometric
models. Econometric Theory, 1-48.

\item Costantini, P. and Fontanella, F. (1990): Shape-preserving bivariate
interpolation. SIAM Journal on Numerical Analysis 27(2), 488-506.

\item Dette, H., Hoderlein, S. and Neumeyer, N. (2016): Testing multivariate
economic restrictions using quantiles: the example of Slutsky negative
semidefiniteness. Journal of Econometrics 191(1), 129-144.

\item Ghosal, S., Sen, A. and Van Der Vaart, A.W. (2000): Testing
monotonicity of regression. The Annals of Statistics 28(4), 1054-1082.

\item Hall, P. and Heckman, N.E. (2000): Testing for monotonicity of a
regression mean by calibrating for linear functions. The Annals of
Statistics 28(1), 20-39.

\item Han, A.K. (1987): Non-parametric analysis of a generalized regression
model: the maximum rank correlation estimator. Journal of Econometrics,
35(2-3), 303-316.

\item Hausman, J.A. and Newey, W.K. (2016): Individual heterogeneity and
average welfare. Econometrica, 84(3), 1225-1248.

\item Hoderlein, S. (2011): How many consumers are rational?, Journal of
Econometrics 164(2), 294-309.

\item Hoderlein, S. and Stoye, J. (2014): Revealed preferences in a
heterogeneous population. Review of Economics and Statistics, 96(2), 197-213.

\item Imbens, G.W. and Newey, W.K. (2009): Identification and estimation of
triangular simultaneous equations models without additivity. Econometrica,
77(5), 1481-1512.

\item Kitamura, Y. and Stoye, J. (2016): Nonparametric analysis of random
utility models, Econometrica, 86(6), 1883-1909.

\item Klein, R.W. and Spady, R.H. (1993): An efficient semiparametric
estimator for binary response models. Econometrica, 387-421.

\item Lee, Y.Y. and Bhattacharya, D. (2019): Applied welfare analysis for
discrete choice with interval-data on income, Journal of econometrics 211,
no. 2, 361-387.

\item Lewbel, A. (2001): Demand Systems with and without Errors. American
Economic Review, 611-618.

\item McFadden, D. (1973): Conditional logit analysis of qualitative choice
behavior.

\item McFadden, D. and Richter, M.K. (1990): Stochastic rationality and
revealed stochastic preference. Preferences, Uncertainty, and Optimality,
Essays in Honor of Leo Hurwicz, Westview Press, 161-186.

\item McFadden, D. (2005): Revealed Stochastic Preference: A Synthesis.
Economic Theory 26(2): 245--264.

\item Manski, C.F. (1975): Maximum score estimation of the stochastic
utility model of choice. Journal of econometrics, 3(3), 205-228.

\item Matzkin, R.L. (1992): Nonparametric and distribution-free estimation
of the binary threshold crossing and the binary choice models. Econometrica,
239-270.

\item Pfeiffer, P.E. (1990): Probability for applications. Springer Science
\& Business Media.

\item Train, K.E. (2009): Discrete choice methods with simulation. Cambridge
University Press.\pagebreak 
\end{enumerate}

\begin{center}
\textbf{Online Appendix\smallskip }
\end{center}

\textbf{Abstract}: This online appendix contains: (i) the construction of
the continuous extension of the choice probability function to a domain
containing $\Omega $, as mentioned in Footnote 11 in the proof of Theorem 1,
and (ii) a version of Theorem 1 (called Theorem 2) with proof that does not
require the limit conditions C/C' of Theorem 1, but involves a slight
strengthening of the continuity conditions B/B'.

\begin{center}
\textbf{1. Construction of Continuous Extension of Choice Probability
Function}
\end{center}

In the proof of Theorem 1, the definition of $q^{-1}\left( \cdot
,a_{1}\right) $ in (\ref{4}) implicitly assumes that $\Omega _{0}\left(
a_{1}\right) $ equals (or contains) $[y_{L}\left( a_{1}\right) ,y_{H}\left(
a_{1}\right) ]$. If however the support of price and income are discrete,
then $\Omega _{0}\left( a_{1}\right) $ can be a strict subset of $%
[y_{L}\left( a_{1}\right) ,y_{H}\left( a_{1}\right) ]$. Then $q\left( \cdot
,\cdot \right) $ is not defined at the points `in between' the points of
support, and therefore, $q^{-1}\left( \cdot ,a_{1}\right) $ in (\ref{4}) is
not well-defined. To cover this case, one can extend $q\left( \cdot ,\cdot
\right) $ to a continuous function $q^{c}\left( \cdot ,\cdot \right) $
defined on a rectangle $\Omega ^{c}$ containing $\Omega $ such that (i) $%
q^{c}\left( \cdot ,\cdot \right) $ equals $q\left( \cdot ,\cdot \right) $ on 
$\Omega $, (ii) $q^{c}\left( \cdot ,\cdot \right) $ satisfies the same shape
restrictions on $\Omega ^{c}$ that are satisfied by $q\left( \cdot ,\cdot
\right) $ on $\Omega $, and (iii) $q^{c}\left( \cdot ,\cdot \right) $
satisfies the limit conditions C of Theorem 1. The proof of Theorem 1 then
holds with $\Omega $, $\Omega _{0}\left( \cdot \right) $ and $q\left( \cdot
,\cdot \right) $ equalling their corresponding extensions in the case where $%
\left( P,Y\right) $ have discrete support. Here we provide an explicit
construction that achieves this extension.\footnote{%
Alternatively, one can construct $q^{c}\left( \cdot ,\cdot \right) $ as a
smooth, tensor-product polynomial spline with coefficients chosen to satisfy
the shape restrictions and a high enough degree to guarantee that $%
q^{c}\left( \cdot ,\cdot \right) $ passes through the interpolating points $%
\left\{ y^{j},y^{j}-p^{j},q\left( y^{j},y^{j}-p^{j}\right) :\left(
y^{j},y^{j}-p^{j}\right) \in \Omega \right\} $, along the lines of
Costantini and Fontanella 1990.}

Suppose the support of $\left( P,Y\right) $ is the discrete set $\bar{\Omega}%
=\left\{ p_{1},...,p_{M}\right\} \times \left\{ y_{1},...,y_{N}\right\} $,
with $p_{1}<p_{2}<...<p_{M}$ and $y_{1}<y_{2}<...<y_{N}$. Suppose the choice
probability $q\left( y,y-p\right) $, which is defined for $\left( p,y\right)
\in \bar{\Omega}$, satisfies the shape constraints (A) of Theorem 1, i.e. $%
q\left( \cdot ,\cdot \right) $ is non-increasing in the first and
non-decreasing in the second argument. We want to construct an extension of $%
q\left( \cdot ,\cdot \right) $, denoted by $q^{c}\left( y,y-p\right) $,
which is (i) defined for all $\left( y,y-p\right) $ with $p_{1}\leq p\leq
p_{M}$ and $y_{1}\leq y\leq y_{N}$, (ii) equals $q\left( y,y-p\right) $ for $%
\left( p,y\right) \in \bar{\Omega}$, and (iii) satisfies all three
conditions A, B, C of Theorem 1. The construction proceeds in three steps.

\textbf{Step 1:} First, we extend $q\left( \cdot ,\cdot \right) $ to the
rectangular grid%
\begin{equation*}
T=\left\{ y_{1},...,y_{N}\right\} \times \cup _{j=1}^{N}\cup
_{k=1}^{M}\left\{ y_{j}-p_{k}\right\} \text{.}
\end{equation*}%
To do this, define $\tilde{q}\left( \cdot ,\cdot \right) :T\rightarrow \left[
0,1\right] $ as:%
\begin{equation}
\tilde{q}\left( y,y-p\right) =\lambda \bar{L}\left( y,y-p\right) +\left(
1-\lambda \right) \bar{U}\left( y,y-p\right)  \label{C}
\end{equation}%
where $\lambda \in \left[ 0,1\right] $ is arbitrary, and for any $\left(
y,y-p\right) \in T,$

\begin{eqnarray*}
\bar{L}\left( y,y-p\right) &=&\left\{ 
\begin{array}{l}
\sup_{\left( p^{\prime },y^{\prime }\right) \in \bar{\Omega}:\text{ }%
y^{\prime }\geq y,\text{ }y^{\prime }-p^{\prime }\leq y-p}q\left( y^{\prime
},y^{\prime }-p^{\prime }\right) \text{, if }\left\{ \left( p^{\prime
},y^{\prime }\right) \in \bar{\Omega}:\text{ }y^{\prime }\geq y,\text{ }%
y^{\prime }-p^{\prime }\leq y-p\right\} \neq \phi \\ 
0\text{, if }\left\{ \left( p^{\prime },y^{\prime }\right) \in \bar{\Omega}:%
\text{ }y^{\prime }\geq y,\text{ }y^{\prime }-p^{\prime }\leq y-p\right\}
=\phi%
\end{array}%
\right. \\
\bar{U}\left( y,y-p\right) &=&\left\{ 
\begin{array}{l}
\inf_{\left( p^{\prime },y^{\prime }\right) \in \bar{\Omega}:\text{ }%
y^{\prime }\leq y,\text{ }y^{\prime }-p^{\prime }\geq y-p}q\left( y^{\prime
},y^{\prime }-p^{\prime }\right) \text{, if }\left\{ \left( p^{\prime
},y^{\prime }\right) \in \bar{\Omega}:\text{ }y^{\prime }\leq y,\text{ }%
y^{\prime }-p^{\prime }\geq y-p\right\} \neq \phi \\ 
1\text{, if }\left\{ \left( p^{\prime },y^{\prime }\right) \in \bar{\Omega}:%
\text{ }y^{\prime }\leq y,\text{ }y^{\prime }-p^{\prime }\geq y-p\right\}
=\phi%
\end{array}%
\right.
\end{eqnarray*}

Note that $\tilde{q}\left( \cdot ,\cdot \right) $, which is well defined on
all of $T$, satisfies the shape constraints (A) of Theorem 1. This is
because the set $\left\{ \left( p^{\prime },y^{\prime }\right) \in \bar{%
\Omega}:\text{ }y^{\prime }\geq y,\text{ }y^{\prime }-p^{\prime }\leq
y-p\right\} $ is decreasing in $y$ for fixed $y-p$, and increasing in $y-p$
for fixed $y$, so $\bar{L}\left( \cdot ,\cdot \right) $ is decreasing in the
first and increasing in the second argument; an analogous argument works for 
$\bar{U}\left( \cdot ,\cdot \right) $. Furthermore, if $\left( p,y\right)
\in \bar{\Omega}$, then%
\begin{eqnarray*}
\left( p,y\right) &\in &\left\{ \left( p^{\prime },y^{\prime }\right) \in 
\bar{\Omega}:\text{ }y^{\prime }\geq y,\text{ }y^{\prime }-p^{\prime }\leq
y-p\right\} \text{,} \\
\left( p,y\right) &\in &\left\{ \left( p^{\prime },y^{\prime }\right) \in 
\bar{\Omega}:\text{ }y^{\prime }\leq y,\text{ }y^{\prime }-p^{\prime }\geq
y-p\right\} \text{,}
\end{eqnarray*}%
whence the shape restrictions on $q\left( \cdot ,\cdot \right) $ imply that $%
\bar{L}\left( y,y-p\right) =q\left( y,y-p\right) =\bar{U}\left( y,y-p\right) 
$, and hence $\tilde{q}\left( y,y-p\right) =q\left( y,y-p\right) $. Note,
however, that $\tilde{q}\left( \cdot ,\cdot \right) $ does not satisfy the
continuity condition (B) and the limit conditions (C) of Theorem 1.\smallskip

\textbf{Step 2:} The second step is to extend $\tilde{q}\left( \cdot ,\cdot
\right) $ to a \emph{continuous} function $q^{c}\left( \cdot ,\cdot \right) $
on the entire rectangle $\left[ y_{1},y_{N}\right] \times \left[ y_{1}-p_{M},%
\text{ }y_{N}-p_{1}\right] $, satisfying the shape constraints (A) of
theorem 1, while also satisfying the interpolation conditions $q^{c}\left(
y,y-p\right) =q\left( y,y-p\right) $ for $\left( p,y\right) \in \bar{\Omega}$%
. This is done using bilinear shape-preserving interpolation as follows.

Recall $y_{1}<y_{2}<...<y_{N}$, and define $w_{1}<w_{2}<...<w_{J}$ with $%
J\leq MN$ to be the ordered values of the set $\left\{
y_{1}-p_{1},...,y_{1}-p_{M},...,y_{N}-p_{1},...,y_{N}-p_{M}\right\} $. We
can have $J<MN$ if for some $\left( j,k\right) \neq \left( l,m\right) $, it
holds that $y_{j}-p_{k}=y_{l}-p_{m}$. For each $i=1,...N-1$, $j=1,...,J-1$,
and for $\left( y,y-p\right) \in \lbrack y_{i},y_{i+1}]\times \lbrack
w_{j},w_{j+1}]$, let%
\begin{eqnarray}
\alpha _{i}\left( y\right) &=&\frac{y-y_{i}}{y_{i+1}-y_{i}}\text{, \ \ }%
\beta _{j}\left( w\right) =\frac{w-w_{j}}{w_{j+1}-w_{j}}\text{,}  \notag \\
q^{c}\left( y,\underset{w}{\underbrace{y-p}}\right) &=&\left( 1-\alpha
_{i}\left( y\right) \right) \times \left( 1-\beta _{j}\left( w\right)
\right) \times \tilde{q}\left( y_{i},w_{j}\right)  \notag \\
&&+\alpha _{i}\left( y\right) \times \left( 1-\beta _{j}\left( w\right)
\right) \times \tilde{q}\left( y_{i+1},w_{j}\right)  \notag \\
&&+\left( 1-\alpha _{i}\left( y\right) \right) \times \beta _{j}\left(
w\right) \times \tilde{q}\left( y_{i},w_{j+1}\right)  \notag \\
&&+\alpha _{i}\left( y\right) \times \beta _{j}\left( w\right) \times \tilde{%
q}\left( y_{i+1},w_{j+1}\right) \text{,}  \label{E}
\end{eqnarray}%
where $\tilde{q}\left( \cdot ,\cdot \right) $ is defined in (\ref{C}).

\textbf{Step 3}: The last step in the construction is to extend $q^{c}\left(
\cdot ,\cdot \right) $ beyond $\left[ y_{1},y_{N}\right] \times \left[
y_{1}-p_{M},\text{ }y_{N}-p_{1}\right] $ to ensure that the limit conditions
(C) of Theorem 1 are satisfied. To do this, choose any pair of real numbers $%
y_{L},y_{H}$ s.t. $y_{L}<y_{1}$ and $y_{H}>y_{N}$. Let 
\begin{equation*}
D=\left[ y_{L},y_{H}\right] \times \left[ y_{1}-p_{M},\text{ }y_{N}-p_{1}%
\right] .
\end{equation*}%
For any $w\in \left[ y_{1}-p_{M},\text{ }y_{N}-p_{1}\right] $, define%
\begin{equation}
q^{c}\left( y,w\right) =\left\{ 
\begin{array}{l}
\frac{y-y_{L}}{y_{1}-y_{L}}\times q^{c}\left( y_{1},w\right) +\frac{y_{1}-y}{%
y_{1}-y_{L}}\text{, if }y\in \left[ y_{L},y_{1}\right] \\ 
\frac{y_{H}-y}{y_{H}-y_{N}+p_{1}}q^{c}\left( y_{N}-p_{1},w\right) \text{, if 
}y\in \left[ y_{N}-p_{1},y_{H}\right]%
\end{array}%
\right.  \label{A}
\end{equation}%
That is for $y\in \left[ y_{L},y_{1}\right] $, $q^{c}\left( y,w\right) $ is
the negatively sloped straight line joining $q^{c}\left( y_{1},w\right) $ to 
$1\equiv q^{c}\left( y_{L},w\right) $, and for $y\in \left[ y_{N}-p_{1},y_{H}%
\right] $, $q^{c}\left( y,w\right) $ is the negatively sloped straight line
joining $q^{c}\left( y_{N}-p_{1},w\right) $ to $0\equiv q^{c}\left(
y_{H},w\right) $.

\textbf{Proof that }$q^{c}\left( \cdot ,\cdot \right) :D\rightarrow \left[
0,1\right] $\textbf{\ equals }$q\left( y,y-p\right) $\textbf{\ for }$\left(
p,y\right) \in \bar{\Omega}$\textbf{\ and satisfies conditions (A), (B) ,(C)
of Theorem 1}: To see the first assertion, observe that at the grid points $%
y=y_{i}$, $y-p=w_{j}$, we get from (\ref{E}) that $\alpha _{i}\left(
y\right) =0=\beta _{j}\left( w\right) $, so that $q^{c}\left( y,w\right) =%
\tilde{q}\left( y_{i},w_{j}\right) $. We have already seen that for $\left(
p,y\right) \in \bar{\Omega}$, $q\left( y,y-p\right) =\tilde{q}\left(
y,y-p\right) $. Now, since $\left( p,y\right) \in \bar{\Omega}$ implies $%
\left( y,y-p\right) \in T$, putting these two conclusions together, we get
that for $\left( p,y\right) \in \bar{\Omega}$, it holds that $q^{c}\left(
y,y-p\right) =\tilde{q}\left( y,y-p\right) =q\left( y,y-p\right) $.

As for the continuity condition (B) of Theorem 1, observe that holding fixed 
$w$, as $y\in \lbrack y_{i},y_{i+1})\nearrow y_{i+1}-$, we have that $\alpha
_{i}\left( y\right) \nearrow 1$ whence from (\ref{E}), it follows that%
\begin{equation}
q^{c}\left( y,w\right) \searrow \left( 1-\beta _{j}\left( w\right) \right)
\times \tilde{q}\left( y_{i+1},w_{j}\right) +\beta _{j}\left( w\right)
\times \tilde{q}\left( y_{i+1},w_{j+1}\right) \text{.}  \label{D}
\end{equation}%
On the other hand, for the same $w$ and for $y\in \lbrack y_{i+1},y_{i+2})$,
we have that $\alpha _{i}\left( y\right) =\frac{y-y_{i+1}}{y_{i+2}-y_{i+1}}$
which at $y=y_{i+1}\in \lbrack y_{i+1},y_{i+2})$ equals $0$, whence from (%
\ref{E}) with $i$ replaced by $i+1$ and $i+1$ replaced by $i+2$, we get%
\begin{equation*}
q^{c}\left( y,w\right) =\left( 1-\beta _{j}\left( w\right) \right) \times 
\tilde{q}\left( y_{i+1},w_{j}\right) +\beta _{j}\left( w\right) \times 
\tilde{q}\left( y_{i+1},w_{j+1}\right)
\end{equation*}%
which equals (\ref{D}). Therefore, for fixed $w$, $\tilde{q}\left(
y,w\right) $ is simply a piecewise linear function of $y$ joined at the
end-points $y_{2},...,y_{N-1}$, and therefore continuous in $y$ for $y\in %
\left[ y_{1},y_{N}\right] $. For $y\in \left[ y_{L},y_{H}\right] \backslash %
\left[ y_{1},y_{N}\right] $, continuity is obvious from (\ref{A}) and the
fact that $\lim_{y\nearrow y_{1}-}q^{c}\left( y,w\right) =q^{c}\left(
y_{1},w\right) =\lim_{y\searrow y_{1}+}q^{c}\left( y,w\right) $ and $%
\lim_{y\nearrow \left( y_{N}-p_{1}\right) -}q^{c}\left( y,w\right)
=q^{c}\left( y_{N}-p_{1},w\right) =\lim_{y\searrow \left( y_{N}-p_{1}\right)
+}q^{c}\left( y,w\right) $. An analogous argument shows that $q^{c}\left(
y,w\right) $ is also continuous in $w$ for fixed $y$ (this property is not
needed to prove Theorem 1 but is used in Theorem 2, the alternative version
of Theorem 1 without the limiting condition, which appears below).

The limiting conditions (C) of Theorem 1 are satisfied, since (\ref{A})
implies that $q^{c}\left( y_{L},w\right) =1$ and $q^{c}\left( y_{H},w\right)
=0$ for each $w\in \left[ y_{1}-p_{M},\text{ }y_{N}-p_{1}\right] $.

Finally, to see that the shape restrictions (A) of Theorem 1 hold on $\left[
y_{1},y_{N}\right] \times \left[ y_{1}-p_{M},\text{ }y_{N}-p_{1}\right] $,
note from (\ref{E}) that the coefficient of $y$ in $q^{c}\left( y,w\right) $
equals%
\begin{equation*}
\underset{\geq 0}{\underbrace{\frac{1}{y_{i+1}-y_{i}}}}\times \left\{ 
\begin{array}{c}
\underset{\geq 0}{\underbrace{\left( 1-\beta _{j}\left( w\right) \right) }}%
\times \underset{\leq 0\text{, since }y_{i}\leq y_{i+1}}{\underbrace{\left[ 
\tilde{q}\left( y_{i+1},w_{j}\right) -\tilde{q}\left( y_{i},w_{j}\right) %
\right] }} \\ 
\underset{\leq 0}{\underbrace{-\beta _{j}\left( w\right) }}\times \left[ 
\underset{\geq 0\text{, since }y_{i}\leq y_{i+1}}{\underbrace{\tilde{q}%
\left( y_{i},w_{j+1}\right) -\tilde{q}\left( y_{i+1},w_{j+1}\right) }}\right]%
\end{array}%
\right\} \leq 0\text{.}
\end{equation*}%
Similarly, the coefficient of $w$ in $q^{c}\left( y,w\right) $ equals%
\begin{equation*}
\underset{\geq 0}{\underbrace{\frac{1}{w_{j+1}-w_{j}}}}\times \left\{ 
\begin{array}{c}
\underset{\geq 0}{\underbrace{\left( 1-\alpha _{i}\left( y\right) \right) }}%
\times \underset{\geq 0\text{, since }w_{j}\leq w_{j+1}}{\underbrace{\left[ 
\tilde{q}\left( y_{i},w_{j+1}\right) -\tilde{q}\left( y_{i},w_{j}\right) %
\right] }} \\ 
\underset{\geq 0}{\underbrace{+\alpha _{i}\left( y\right) }}\times \left[ 
\underset{\geq 0\text{, since }w_{j}\leq w_{j+1}}{\underbrace{\tilde{q}%
\left( y_{i+1},w_{j+1}\right) -\tilde{q}\left( y_{i+1},w_{j}\right) }}\right]%
\end{array}%
\right\} \geq 0\text{.}
\end{equation*}%
From (\ref{A}) it follows that the shape restrictions also hold on $\left[
y_{L},y_{1}\right] \times \left[ y_{1}-p_{M},\text{ }y_{N}-p_{1}\right] $
and on $\left[ y_{N},y_{H}\right] \times \left[ y_{1}-p_{M},\text{ }%
y_{N}-p_{1}\right] $, and thus condition (A) of Theorem 1 holds on all of $%
\left[ y_{L},y_{H}\right] \times \left[ y_{1}-p_{M},\text{ }y_{N}-p_{1}%
\right] $.

Thus $q^{c}\left( \cdot ,\cdot \right) $ satisfies all three conditions of
Theorem 1.\medskip

\begin{center}
\textbf{2. Main Result without condition (C/C')}
\end{center}

The following is a version of Theorem 1 that does not require the technical
conditions C and C' of Theorem 1, but involves a slight strengthening of the
technical condition B. The proof of this version is considerably longer than
that of Theorem 1. The proof works by constructing an extension $Q\left(
\cdot ,\cdot \right) $ of $q\left( \cdot ,\cdot \right) $ which satisfies
properties (A)-(C) of Theorem 1 although $q\left( \cdot ,\cdot \right) $
itself does not satisfy property (C).\footnote{%
The case where $\left( P,Y\right) $ have a discrete support is handled in
exactly the same way as in Theorem 1 with two small modifications: (a) Step
3 in the construction immediately above is not required, and (b) continuity
of $q^{c}\left( \cdot ,\cdot \right) $ in the \textit{second} argument is
guaranteed by the construction in Step 2.}

Suppose the support of price $P$ and income $Y$ in the population is $%
[p_{l},p_{u}]\times \left[ y_{l},y_{u}\right] $. Correspondingly, the
support of $Y-P$ is $\Omega _{1}\overset{defn}{=}\left[
y_{l}-p_{u},y_{u}-p_{l}\right] $. Pick any $a_{1}\in \Omega _{1}$.
Corresponding to $Y-P=a_{1}$, the support of $Y=a_{1}+P$ is therefore 
\begin{equation*}
\Omega _{0}\left( a_{1}\right) \overset{defn}{=}\left[ \underset{L\left(
a_{1}\right) }{\underbrace{\max \left\{ p_{l}+a_{1},y_{l}\right\} }},%
\underset{U\left( a_{1}\right) }{\underbrace{\min \left\{
p_{u}+a_{1},y_{u}\right\} }}\right] \text{.}
\end{equation*}%
Note that by definition, $L\left( \cdot \right) $ and $U\left( \cdot \right) 
$ are non-decreasing and continuous. Let $\Omega =\cup _{a_{1}\in \Omega
_{1}}\cup _{a_{0}\in \Omega _{0}\left( a_{1}\right) }\left\{
a_{0},a_{1}\right\} $.\medskip

\begin{theorem}
For binary choice under general heterogeneity, the following two statements
are equivalent:

(I) The choice probabilities $q\left( y,y-p\right) $, defined above, satisfy
that (A) $q\left( \cdot ,y-p\right) $ is non-increasing, and $q\left(
y,\cdot \right) $ is non-decreasing; (B) $q\left( \cdot ,\cdot \right) $ is
continuous.

(II) There exists a pair of utility functions $W_{0}\left( \cdot ,\eta
\right) $ and $W_{1}\left( \cdot ,\eta \right) $, where the first argument
denotes the amount of numeraire, and $\eta $ denotes unobserved
heterogeneity, and a distribution $G\left( \cdot \right) $ of $\eta $ such
that for any $\left( y-p\right) \in \Omega _{1}$ and correspondingly $y\in
\Omega _{0}\left( y-p\right) $,%
\begin{equation*}
q\left( y,y-p\right) =\int 1\left\{ W_{0}\left( y,\eta \right) \leq
W_{1}\left( y-p,\eta \right) \right\} dG\left( \eta \right) \text{,}
\end{equation*}%
where (A') for each fixed $\eta $, $W_{0}\left( \cdot ,\eta \right) $ and $%
W_{1}\left( \cdot ,\eta \right) $ are non-decreasing; (B') for each fixed $%
\eta $, $W_{0}\left( \cdot ,\eta \right) $ and $W_{1}\left( \cdot ,\eta
\right) $ are continuous, and for any $\left( a_{0},a_{1}\right) \in \Omega $%
, it holds that $\int 1\left\{ W_{0}\left( a_{0},\eta \right) \leq
W_{1}\left( a_{1},\eta \right) \right\} dG\left( \eta \right) $ is
continuous in $\left( a_{0},a_{1}\right) $.\smallskip
\end{theorem}

\textbf{Discussion of assumptions}: Relative to Theorem 1, conditions (C/C')
are omitted, and condition (B/B') is strengthened to continuity in both
arguments. Note that under monotonicity in any one argument, the joint
continuity of $q\left( \cdot ,\cdot \right) $ is equivalent to coordinate
wise continuity c.f. Kruse and Deely 1969.\smallskip

To prove Theorem 2, we will utilize several lemmas.

\begin{lemma}[Apostol, 1974, Ex 4.19]
Suppose $r\left( \cdot \right) :\left[ c,b\right] \rightarrow \mathbb{R}$,
is continuous on $\left[ c,b\right] $. For $x\in \left[ c,b\right] $, define 
$g\left( x\right) =\sup \left\{ r\left( z\right) :x\leq z\leq b\right\} $,
and $h\left( x\right) =\sup \left\{ r\left( z\right) :c\leq z\leq x\right\} $%
. Then $g\left( \cdot \right) $ and $h\left( \cdot \right) $ are continuous
on $\left[ c,b\right] $.
\end{lemma}

\begin{proof}[Proof of Lemma 2]
Fix any $x\in \left[ c,a_{1}\right] $.

First, suppose $g\left( x\right) >r\left( x\right) $. Choose $\varepsilon
=g\left( x\right) -r\left( x\right) >0$. Now by continuity of $r\left( \cdot
\right) $, there must exist $\delta >0$ s.t. for any $z\in \lbrack x-\delta
,x+\delta ]$, we have that $r\left( z\right) <r\left( x\right) +\varepsilon
=r\left( x\right) +g\left( x\right) -r\left( x\right) =g\left( x\right) $.
Therefore, $\sup \left\{ r\left( z\right) :x-\delta \leq z\leq x+\delta
\right\} <g\left( x\right) $. Therefore, $g\left( x-\delta \right) =g\left(
x\right) =g\left( x+\delta \right) $, implying continuity of $g\left( \cdot
\right) $ at $x$.

Next, suppose the sup is at $x$, i.e. $g\left( x\right) =r\left( x\right) $.
By continuity, for any $\varepsilon >0$, there exists $\delta >0$, s.t. for
all $u\in \left[ x-\delta ,x+\delta \right] $, we have that $r\left(
x\right) +\varepsilon \geq r\left( u\right) \geq r\left( x\right)
-\varepsilon $. For $u\in \left[ x,x+\delta \right] $, $g\left( u\right)
=\sup \left\{ r\left( z\right) :u\leq z\leq a_{1}\right\} \geq r\left(
u\right) \geq r\left( x\right) -\varepsilon =g\left( x\right) -\varepsilon $%
, since $g\left( x\right) =r\left( x\right) $, by assumption. But $g\left(
u\right) \leq g\left( x\right) $ by definition. Therefore, for all $u\in %
\left[ x,x+\delta \right] $, we have that $g\left( x\right) \geq g\left(
u\right) >g\left( x\right) -\varepsilon $. Next, for all $u\in \left[
x-\delta ,x\right] $, $r\left( u\right) \leq r\left( x\right) +\varepsilon
=g\left( x\right) +\varepsilon $ implying%
\begin{eqnarray*}
g\left( u\right) &=&\sup \left\{ r\left( z\right) :u\leq z\leq a_{1}\right\}
\\
&\leq &\sup \left\{ r\left( z\right) :x-\delta \leq z\leq a_{1}\right\} \\
&=&\max \left\{ \underset{\leq g\left( x\right) +\varepsilon }{\underbrace{%
\sup \left\{ r\left( z\right) :x-\delta \leq z\leq x\right\} }},\underset{%
g\left( x\right) }{\underbrace{\sup \left\{ r\left( z\right) :x\leq z\leq
a_{1}\right\} }}\right\} \\
&\leq &g\left( x\right) +\varepsilon .
\end{eqnarray*}%
Thus for all $u\in \left[ x-\delta ,x+\delta \right] $, we have that $%
g\left( x\right) +\varepsilon \geq g\left( u\right) >g\left( x\right)
-\varepsilon $. Therefore, $g\left( \cdot \right) $ is continuous at $x$.

An exactly similar proof works for $h\left( x\right) =\sup \left\{ r\left(
z\right) :c\leq z\leq x\right\} $.\smallskip
\end{proof}

\begin{lemma}[Taylor, 1955, Chap 15.7, Theorem VII]
Suppose the function $f:\mathbb{R}^{2}\rightarrow \mathbb{R}$ is continuous,
and the function $g\left( \cdot \right) :\mathbb{R}\rightarrow \mathbb{R}$
is continuous w.r.t. the $L1$-norm. Then the function $h:\mathbb{R}%
\rightarrow \mathbb{R}$ defined as $h\left( x\right) =f\left( g\left(
x\right) ,x\right) $ is continuous on $\mathbb{R}$.
\end{lemma}

\begin{proof}[Proof of Lemma 3]
Pick any $x_{0}\in \mathbb{R}$, and $\varepsilon >0$. Continuity of $f\left(
\cdot ,\cdot \right) $ implies that there exists $\delta >0$ s.t. $%
\left\vert f\left( g\left( x\right) ,x\right) -f\left( g\left( x_{0}\right)
,x_{0}\right) \right\vert \leq \varepsilon $, whenever $\left\Vert \left(
g\left( x\right) ,x\right) -\left( g\left( x_{0}\right) ,x_{0}\right)
\right\Vert \leq \delta $. Now, continuity of $g\left( \cdot \right) $
implies that given the above $\delta >0$, there exists $\delta _{1}>0$ s.t. $%
\left\vert g\left( x\right) -g\left( x_{0}\right) \right\vert \leq \delta /2$
whenever $\left\vert x-x_{0}\right\vert \leq \delta _{1}$. Choose $\delta
^{\ast }=\min \left\{ \delta /2,\delta _{1}\right\} $. Then whenever $%
\left\vert x-x_{0}\right\vert \leq \delta ^{\ast }$, we have that $%
\left\vert g\left( x\right) -g\left( x_{0}\right) \right\vert \leq \delta /2$
and $\left\vert x-x_{0}\right\vert \leq \delta /2$, and thus $\left\Vert
\left( g\left( x\right) ,x\right) -\left( g\left( x_{0}\right) ,x_{0}\right)
\right\Vert =\left\vert g\left( x\right) -g\left( x_{0}\right) \right\vert
+\left\vert x-x_{0}\right\vert \leq \delta $, and therefore,%
\begin{equation*}
\left\vert h\left( x\right) -h\left( x_{0}\right) \right\vert =\left\vert
f\left( g\left( x\right) ,x\right) -f\left( g\left( x_{0}\right)
,x_{0}\right) \right\vert <\varepsilon \text{.}
\end{equation*}
\end{proof}

\textbf{Construction}: The following construction will be used to prove the
theorem. Pick $a_{1}\in \Omega _{1}$. Recall the definitions $L\left(
a_{1}\right) \equiv \max \left\{ p_{l}+a_{1},y_{l}\right\} ,$ and $U\left(
a_{1}\right) \equiv \min \left\{ p_{u}+a_{1},y_{u}\right\} $. Let $a_{0L}$, $%
a_{0H}$ be any pair of real numbers satisfying $a_{0L}<y_{l}$ and $%
a_{0H}>y_{u}$. For any $a_{0}<L\left( a_{1}\right) $ and $a_{0}>U\left(
a_{1}\right) $, respectively, define%
\begin{eqnarray*}
H\left( a_{0},a_{1}\right) &=&\sup \left\{ q\left( L\left( x\right)
,x\right) :L\left( x\right) \in \left[ a_{0},L\left( a_{1}\right) \right]
\right\} \text{,} \\
h\left( a_{0},a_{1}\right) &=&\inf \left\{ q\left( U\left( x\right)
,x\right) :U\left( x\right) \in \left[ U\left( a_{1}\right) ,a_{0}\right]
\right\} \text{.}
\end{eqnarray*}%
Note that as $a_{0}$ decreases with $a_{1}$ fixed, or $a_{1}$ increases with 
$a_{0}$ fixed, the set $\left[ a_{0},L\left( a_{1}\right) \right] $ expands,
and therefore the sup over it weakly increases; thus $H\left( \cdot
,a_{1}\right) $ is non-increasing and $H\left( a_{0},\cdot \right) $ is
non-decreasing. Similarly, $h\left( \cdot ,a_{1}\right) $ is non-increasing
and $h\left( a_{0},\cdot \right) $ is non-decreasing. Now, define the
function $Q\left( \cdot ,\cdot \right) :\left[ a_{0L},a_{0H}\right] \mathbb{%
\rightarrow }\left[ 0,1\right] $ as follows. For any $a_{1}\in \Omega _{1}$,%
\begin{equation}
Q\left( a_{0},a_{1}\right) =\left\{ 
\begin{array}{l}
H\left( y_{l},a_{1}\right) +\left( 1-H\left( y_{l},a_{1}\right) \right) 
\frac{y_{l}-a_{0}}{y_{l}-a_{0L}}\text{, if }a_{0L}\leq a_{0}<y_{l}\text{,}
\\ 
H\left( a_{0},a_{1}\right) \text{, if }y_{l}\leq a_{0}<L\left( a_{1}\right) 
\text{,} \\ 
q\left( a_{0},a_{1}\right) \text{, if }a_{0}\in \left[ L\left( a_{1}\right)
,U\left( a_{1}\right) \right] \text{,} \\ 
h\left( a_{0},a_{1}\right) \text{, if }U\left( a_{1}\right) <a_{0}\leq y_{u}%
\text{,} \\ 
\frac{a_{0H}-a_{0}}{a_{0H}-y_{u}}h\left( y_{u},a_{1}\right) \text{, if }%
y_{u}<a_{0}\leq a_{0H}\text{.}%
\end{array}%
\right.  \label{YY}
\end{equation}

\begin{claim}
Suppose $q\left( \cdot ,\cdot \right) $ satisfies (A) and (B) of Theorem 2.
Then the function $Q\left( \cdot ,\cdot \right) $ defined in (\ref{YY})
satisfies the following properties:

(1) $Q\left( \cdot ,a_{1}\right) $ is non-increasing, and $Q\left(
a_{0},\cdot \right) $ is non-decreasing for all $\left( a_{0},a_{1}\right)
\in \left[ a_{0L},a_{0H}\right] \times \Omega _{1}$

(2) $Q\left( \cdot ,\cdot \right) $ is continuous in each argument, holding
the other argument fixed.

(3) For any $a_{1}\in \Omega _{1}$, there exist real numbers $a_{0L}$ and $%
a_{0H}$ such that $\lim_{a_{0}\searrow a_{0L}}Q\left( a_{0},a_{1}\right) =1$
and $\lim_{a_{0}\nearrow a_{0H}}Q\left( a_{0},a_{1}\right) =0$.
\end{claim}

\begin{proof}
Property (3) is obvious because $Q\left( a_{0L},a_{1}\right) =1$ and $%
Q\left( a_{0H},a_{1}\right) =0$, by construction. To show (1) and (2), fix $%
a_{1}\in \Omega _{1}$. Since $q\left( \cdot ,\cdot \right) $ satisfies (A)
and (B) on $a_{0}\in \left[ L\left( a_{1}\right) ,U\left( a_{1}\right) %
\right] $, we only need to establish the properties over the range $%
a_{0}<L\left( a_{1}\right) $ and $a_{0}>U\left( a_{1}\right) $.

\textbf{Property (1)}: First, we show that the shape restrictions hold for $%
Q\left( \cdot ,\cdot \right) $. We have already noted that $H\left( \cdot
,a_{1}\right) $ and $h\left( \cdot ,a_{1}\right) $ are both non-increasing;
further since $H\left( y_{l},a_{1}\right) \leq 1$ and $h\left(
y_{u},a_{1}\right) \geq 0$, we have that $H\left( y_{l},a_{1}\right) +\left(
1-H\left( y_{l},a_{1}\right) \right) \frac{y_{l}-a_{0}}{y_{l}-a_{0L}}$ is
non-increasing in $a_{0}$ for $a_{0L}\leq a_{0}<y_{l}$, and $\frac{%
a_{0H}-a_{0}}{a_{0H}-y_{u}}h\left( y_{u},a_{1}\right) $ is non-increasing in 
$a_{0}$ for $y_{u}<a_{0}\leq a_{0H}$. Thus $Q\left( a_{0},a_{1}\right) $ is
non-increasing in $a_{0}$ for all $a_{0}<L\left( a_{1}\right) $ and $%
a_{0}>U\left( a_{1}\right) $.

Next, pick $a_{0}\in \left[ a_{0L},a_{0H}\right] $, and consider
monotonicity of $Q\left( a_{0},\cdot \right) $. Let $a_{1}^{1},a_{1}^{2}\in
\Omega _{1}$ with $a_{1}^{1}<a_{1}^{2}$, implying $L\left( a_{1}^{1}\right)
\leq L\left( a_{1}^{2}\right) $ and $U\left( a_{1}^{1}\right) \leq U\left(
a_{1}^{2}\right) $. Now there are 10 cases to consider, labelled (\textbf{a}%
)-(\textbf{j}) below, depending on the ordering of $L\left( a_{1}^{2}\right) 
$ and $U\left( a_{1}^{1}\right) $, and where $a_{0}$ lies. Case (\textbf{a}) 
$a_{0L}\leq a_{0}<y_{l}$, then%
\begin{eqnarray*}
Q\left( a_{0},a_{1}^{1}\right) &=&H\left( a_{0},a_{1}^{1}\right) \\
&=&\frac{y_{l}-a_{0}}{y_{l}-a_{0L}}+H\left( y_{l},a_{1}^{1}\right) \frac{%
a_{0}-a_{0L}}{y_{l}-a_{0L}} \\
&\leq &\frac{y_{l}-a_{0}}{y_{l}-a_{0L}}+H\left( y_{l},a_{1}^{2}\right) \frac{%
a_{0}-a_{0L}}{y_{l}-a_{0L}}\text{, since }H\left( y_{l},\cdot \right) \text{
nondecreasing} \\
&=&Q\left( a_{0},a_{1}^{2}\right) \text{.}
\end{eqnarray*}%
Case (\textbf{b}) $y_{l}\leq a_{0}\leq L\left( a_{1}^{1}\right) $, i.e. $%
\left[ a_{0},L\left( a_{1}^{1}\right) \right] \sqsubseteq \left[
a_{0},L\left( a_{1}^{2}\right) \right] $, and so $H\left(
a_{0},a_{1}^{1}\right) \leq H\left( a_{0},a_{1}^{2}\right) $, and therefore, 
$Q\left( a_{0},a_{1}^{1}\right) =H\left( a_{0},a_{1}^{1}\right) \leq H\left(
a_{0},a_{1}^{2}\right) =Q\left( a_{0},a_{1}^{2}\right) $. Case (\textbf{c}): 
$y_{u}<a_{0}\leq a_{0H}$, and Case (\textbf{d}) $U\left( a_{1}^{2}\right)
<a_{0}\leq y_{u}$, the proofs are exactly analogous to respectively (\textbf{%
a}) and (\textbf{b}) above.

So we are left with the following cases, where Cases (\textbf{e})-(\textbf{g}%
) correspond to $U\left( a_{1}^{1}\right) <L\left( a_{1}^{2}\right) $, and (%
\textbf{h})-(\textbf{j}) to $U\left( a_{1}^{1}\right) \geq L\left(
a_{1}^{2}\right) $.

For Case (\textbf{e}) $L\left( a_{1}^{1}\right) \leq a_{0}\leq U\left(
a_{1}^{1}\right) <L\left( a_{1}^{2}\right) $, since $L\left(
a_{1}^{1}\right) <a_{0}<L\left( a_{1}^{2}\right) $, by continuity of $%
L\left( \cdot \right) $ and the intermediate value theorem, there exists $%
c\in \left[ a_{1}^{1},a_{1}^{2}\right] $ s.t. $a_{0}=L\left( c\right) $.
Therefore,%
\begin{eqnarray*}
Q\left( a_{0},a_{1}^{1}\right) &=&q\left( a_{0},a_{1}^{1}\right) =q\left(
L\left( c\right) ,a_{1}^{1}\right) \\
&&\overset{(1)}{\leq }q\left( L\left( c\right) ,c\right) \\
&&\overset{(2)}{\leq }\sup \left\{ q\left( L\left( x\right) ,x\right)
:L\left( x\right) \in \left[ L\left( c\right) ,L\left( a_{1}^{2}\right) %
\right] \right\} \\
&=&\sup \left\{ q\left( L\left( x\right) ,x\right) :L\left( x\right) \in 
\left[ a_{0},L\left( a_{1}^{2}\right) \right] \right\} \text{, since }%
a_{0}=L\left( c\right) \\
&=&Q\left( a_{0},a_{1}^{2}\right) \text{,}
\end{eqnarray*}%
where $\overset{(1)}{\leq }$ holds because $a_{1}^{1}\leq c$ and condition
(A) of Theorem 1, and $\overset{(2)}{\leq }$ holds by definition of $\sup $.
Next, suppose Case (\textbf{f}) $L\left( a_{1}^{1}\right) \leq U\left(
a_{1}^{1}\right) \leq a_{0}<L\left( a_{1}^{2}\right) \leq U\left(
a_{1}^{2}\right) $, then by continuity of $L\left( \cdot \right) $ and the
intermediate value theorem, there exists $c\in \left[ a_{1}^{1},a_{1}^{2}%
\right] $ s.t. $a_{0}=L\left( c\right) $; and by continuity of $U\left(
\cdot \right) $ and the intermediate value theorem, there exists $d\in \left[
a_{1}^{1},a_{1}^{2}\right] $ s.t. $a_{0}=U\left( d\right) $, with $d\leq c$.
Then 
\begin{eqnarray*}
Q\left( a_{0},a_{1}^{1}\right) &=&\inf \left\{ q\left( U\left( x\right)
,x\right) :U\left( a_{1}^{1}\right) \leq U\left( x\right) \leq a_{0}\right\} 
\text{, by (\ref{YY})} \\
&=&\inf \left\{ q\left( U\left( x\right) ,x\right) :U\left( a_{1}^{1}\right)
\leq U\left( x\right) \leq U\left( d\right) \right\} \text{, by }%
a_{0}=U\left( d\right) \\
&\leq &q\left( U\left( d\right) ,d\right) \text{, since }d\in \left\{
x:U\left( a_{1}^{1}\right) \leq U\left( x\right) \leq U\left( d\right)
\right\} \\
&\leq &q\left( L\left( c\right) ,c\right) \text{, by (Aii) since }U\left(
d\right) =a_{0}=L\left( c\right) \text{ and }d\leq c \\
&\leq &\sup \left\{ q\left( L\left( x\right) ,x\right) :L\left( c\right)
\leq L\left( x\right) \leq L\left( a_{1}^{2}\right) \right\} \text{, since }%
c\in \left\{ x:L\left( c\right) \leq L\left( x\right) \leq L\left(
a_{1}^{2}\right) \right\} \\
&=&\sup \left\{ q\left( L\left( x\right) ,x\right) :a_{0}\leq L\left(
x\right) \leq L\left( a_{1}^{2}\right) \right\} \text{, since }a_{0}=L\left(
c\right) \\
&=&Q\left( a_{0},a_{1}^{2}\right) \text{, by definition (\ref{YY}).}
\end{eqnarray*}

Next, for Case (\textbf{g}) $L\left( a_{1}^{1}\right) \leq U\left(
a_{1}^{1}\right) <L\left( a_{1}^{2}\right) \leq a_{0}\leq U\left(
a_{1}^{2}\right) $, using continuity of $U\left( \cdot \right) $ and the
intermediate value theorem, we have $a_{0}=U\left( c\right) $ for some $c\in %
\left[ a_{1}^{1},a_{1}^{2}\right] $ so that%
\begin{eqnarray*}
Q\left( a_{0},a_{1}^{2}\right) &=&Q\left( U\left( c\right) ,a_{1}^{2}\right)
\\
&=&q\left( U\left( c\right) ,a_{1}^{2}\right) \text{, since }a_{0}=U\left(
c\right) \in \left[ L\left( a_{1}^{2}\right) ,U\left( a_{1}^{2}\right) %
\right] \\
&\geq &q\left( U\left( c\right) ,c\right) \text{, since }c\leq a_{1}^{2}%
\text{ and condition (A)} \\
&\geq &\inf \left\{ q\left( U\left( x\right) ,x\right) :U\left(
a_{1}^{1}\right) \leq U\left( x\right) \leq U\left( c\right) \right\} \\
&=&Q\left( U\left( c\right) ,a_{1}^{1}\right) \text{, by (\ref{YY})} \\
&=&Q\left( a_{0},a_{1}^{1}\right) \text{.}
\end{eqnarray*}

Next, consider Case (\textbf{h}) $L\left( a_{1}^{1}\right) \leq a_{0}\leq
L\left( a_{1}^{2}\right) \leq U\left( a_{1}^{1}\right) $. Since $L\left(
a_{1}^{1}\right) \leq a_{0}\leq L\left( a_{1}^{2}\right) $, by continuity
and the intermediate value theorem, we have that $a_{0}=L\left( c\right) $
for some $c\in \left[ a_{1}^{1},a_{1}^{2}\right] $, whence we have%
\begin{eqnarray*}
Q\left( a_{0},a_{1}^{1}\right) &=&q\left( a_{0},a_{1}^{1}\right) =q\left(
L\left( c\right) ,a_{1}^{1}\right) \\
&\leq &q\left( L\left( c\right) ,c\right) \text{, since }c\geq a_{1}^{1} \\
&\leq &\sup \left\{ q\left( L\left( x\right) ,x\right) :L\left( c\right)
\leq L\left( x\right) \leq L\left( a_{1}^{2}\right) \right\} \\
&=&Q\left( L\left( c\right) ,a_{1}^{2}\right) \\
&=&Q\left( a_{0},a_{1}^{2}\right) \text{.}
\end{eqnarray*}

Next, if Case (\textbf{i}) $L\left( a_{1}^{1}\right) \leq L\left(
a_{1}^{2}\right) \leq a_{0}\leq U\left( a_{1}^{1}\right) $, we have that $%
Q\left( a_{0},a_{1}^{1}\right) =q\left( a_{0},a_{1}^{1}\right) \leq q\left(
a_{0},a_{1}^{2}\right) =Q\left( a_{0},a_{1}^{2}\right) $.

Finally, for the Case (\textbf{j}) $L\left( a_{1}^{1}\right) \leq L\left(
a_{1}^{2}\right) \leq U\left( a_{1}^{1}\right) \leq a_{0}\leq U\left(
a_{1}^{2}\right) $, \ the same argument as in (\textbf{g}) applies.

This establishes the requisite shape restrictions, i.e. Property (1).

\textbf{Property (2)}: First, consider continuity of $Q\left( \cdot
,a_{1}\right) $. Note that $H\left( y_{l},a_{1}\right) +\left( 1-H\left(
y_{l},a_{1}\right) \right) \frac{y_{l}-a_{0}}{y_{l}-a_{0L}}$ is obviously
continuous at $a_{0}$ for $a_{0L}\leq a_{0}<y_{l}$; next, at $a_{0}=y_{l}$, $%
Q\left( a_{0},a_{1}\right) =H\left( y_{l},a_{1}\right) +\left( 1-H\left(
y_{l},a_{1}\right) \right) \frac{y_{l}-y_{l}}{y_{l}-a_{0L}}=H\left(
y_{l},a_{1}\right) $, while at $a_{0}=L\left( a_{1}\right) >y_{l}$,%
\begin{equation*}
Q\left( a_{0},a_{1}\right) =\sup \left\{ q\left( L\left( x\right) ,x\right)
:L\left( x\right) \in \left[ L\left( a_{1}\right) ,L\left( a_{1}\right) %
\right] \right\} =q\left( L\left( a_{1}\right) ,a_{1}\right) \text{,}
\end{equation*}%
and thus $Q\left( \cdot ,a_{1}\right) $ is continuous at $a_{0}=y_{l}$ and
at $a_{0}=L\left( a_{1}\right) $. Finally, if $a_{0}\in (y_{l},L\left(
a_{1}\right) )$, then we can have $L\left( x\right) \in \left[ a_{0},L\left(
a_{1}\right) \right] $ only if $L\left( x\right) >y_{l}$ in which case $%
L\left( x\right) =x+p_{l}$ and thus $q\left( L\left( x\right) ,x\right)
=q\left( x+p_{l},x\right) $ implying%
\begin{eqnarray}
Q\left( a_{0},a_{1}\right) &=&\sup \left\{ q\left( L\left( x\right)
,x\right) :a_{0}\leq L\left( x\right) \leq L\left( a_{1}\right) \right\} 
\notag \\
&=&\sup \left\{ q\left( x+p_{l},x\right) :x+p_{l}\in \left[ a_{0},L\left(
a_{1}\right) \right] \right\}  \notag \\
&=&\sup \left\{ q\left( x+p_{l},x\right) :x\in \left[ a_{0}-p_{l},L\left(
a_{1}\right) -p_{l}\right] \right\} \text{.}  \label{XXX}
\end{eqnarray}%
By Lemma 3, $q\left( x+p_{l},x\right) $ is continuous in $x$, and therefore,
by Lemma 2, $Q\left( a_{0},a_{1}\right) $ is continuous in $a_{0}$ for fixed 
$a_{1}$. Thus we have that $Q\left( \cdot ,a_{1}\right) $ is continuous on
all of $\left[ a_{0L},U\left( a_{1}\right) \right] $. An exactly analogous
argument works for $a_{0}>U\left( a_{1}\right) $.

Finally, consider continuity in $a_{1}$ for fixed $a_{0}$. If (a) $a_{1}\leq
y_{l}-p_{l}$, then $L\left( a_{1}\right) =y_{l}$, and therefore,%
\begin{equation}
H\left( a_{0},a_{1}\right) =\sup \left\{ q\left( L\left( x\right) ,x\right)
:L\left( x\right) \in \left[ a_{0},y_{l}\right] \right\}  \label{10'}
\end{equation}%
which does not depend on $a_{1}$ and therefore trivially continuous in $%
a_{1} $. So consider (b) $a_{1}>y_{l}-p_{l}$, so that $L\left( a_{1}\right)
=a_{1}+p_{l}$. Therefore, at $a_{0}=y_{l}$, $H\left( a_{0},a_{1}\right)
=H\left( y_{l},a_{1}\right) $ equals%
\begin{eqnarray}
&&\sup \left\{ q\left( L\left( x\right) ,x\right) :a_{0}\leq L\left(
x\right) \leq L\left( a_{1}\right) \right\}  \label{10} \\
&=&\sup \left\{ q\left( L\left( x\right) ,x\right) :L\left( x\right) \in 
\left[ y_{l},a_{1}+p_{l}\right] \right\}  \notag \\
&&\overset{(2)}{=}\sup \left\{ q\left( L\left( x\right) ,x\right) :x\in %
\left[ y_{l}-p_{u},a_{1}\right] \right\} \text{.}  \notag
\end{eqnarray}%
The last equality $\overset{(2)}{=}$ follows because $\underset{=\max
\left\{ p_{l}+x,y_{l}\right\} }{\underbrace{L\left( x\right) }}\in \left[
y_{l},a_{1}+p_{l}\right] $ if and only if $x\in \left[ y_{l}-p_{u},a_{1}%
\right] $. Now, since $L\left( \cdot \right) $ is continuous, and so is $%
q\left( \cdot ,\cdot \right) $, the function $x\mapsto q\left( L\left(
x\right) ,x\right) $ is continuous in $x$ (see Lemma 3 above), and
therefore, it follows from Lemma 2 that $\sup \left\{ q\left( L\left(
x\right) ,x\right) :x\in \left[ y_{l}-p_{u},a_{1}\right] \right\} $ is
continuous in $a_{1}$. In particular, as $a_{1}\searrow \left(
y_{l}-p_{l}\right) _{+}$, $L\left( a_{1}\right) $ approaches $y_{l}$ and so (%
\ref{10}) tends to (\ref{10'}).

Finally,\textbf{\ }for any $a_{0}>y_{l}$, (recall $a_{1}>y_{l}-p_{l}$, so
that $L\left( a_{1}\right) =a_{1}+p_{l}$), we have that%
\begin{eqnarray*}
H\left( a_{0},a_{1}\right) &=&\sup \left\{ q\left( L\left( x\right)
,x\right) :L\left( x\right) \in \left[ a_{0},a_{1}+p_{l}\right] \right\} \\
&=&\sup \left\{ q\left( L\left( x\right) ,x\right) :x\in \left[
a_{0}-p_{l},a_{1}\right] \right\} \text{,}
\end{eqnarray*}%
which is continuous in $a_{1}$ by Lemma 2 and 3. Exactly analogous arguments
hold for (a') $a_{1}\geq y_{u}-p_{u}$ and (b') $a_{1}<y_{u}-p_{u}$
respectively. Thus, we have that $Q\left( a_{0},\cdot \right) $ is
continuous at each $a_{0}$.
\end{proof}

\begin{lemma}
Suppose the function $Q\left( \cdot ,\cdot \right) :\left[ a_{0L},a_{0H}%
\right] \times \Omega _{1}\sqsubseteq \mathbb{R}^{2}\rightarrow \left[ 0,1%
\right] $ satisfies on its domain that (1) $Q\left( \cdot ,a_{1}\right) $ is
non-increasing, and $Q\left( a_{0},\cdot \right) $ is non-decreasing; (2) $%
Q\left( \cdot ,a_{1}\right) $ is continuous, and (3) for any $a_{1}\in
\Omega _{1}$, $\lim_{a_{0}\searrow a_{0L}}Q\left( a_{0},a_{1}\right) =1$ and 
$\lim_{a_{0}\nearrow a_{0H}}Q\left( a_{0},a_{1}\right) =0$. For any fixed $%
a_{1}\in \Omega _{1}$, define for each $u\in \left[ 0,1\right] $,%
\begin{equation}
Q^{-1}\left( u,a_{1}\right) \overset{def}{=}\sup \left\{ a_{0}\in \left[
a_{0L},a_{0H}\right] :Q\left( a_{0},a_{1}\right) \geq u\right\} \text{.}
\label{4'}
\end{equation}%
Then we must have that $Q\left( Q^{-1}\left( v,a_{1}\right) ,a_{1}\right) =v$%
, for any $v\in \lbrack 0,1]$.
\end{lemma}

\begin{proof}[Proof of Lemma 4]
Since $Q\left( \cdot ,\cdot \right) $ satisfies the same properties as $%
q\left( \cdot ,\cdot \right) $ of Theorem 1 (A)-(C), the proof of this lemma
is identical to the proof of Lemma 1 used to prove Theorem 1.\medskip
\end{proof}

\begin{proof}[Proof of Theorem 2]
That (II) implies (I) is straightforward, since%
\begin{equation*}
q\left( y,y-p\right) =\int 1\left\{ W_{0}\left( y,\eta \right) \leq
W_{1}\left( y-p,\eta \right) \right\} dG\left( \eta \right)
\end{equation*}%
whence (B') implies (B), and (A') implies (A).

We now show that (I) implies (II). To do so, recall the definition of $%
Q^{-1}\left( v,a_{1}\right) $ in (\ref{4'}). Now, consider a random variable 
$V\simeq Uniform\left( 0,1\right) $. Define $W_{0}\left( a_{0},V\right) 
\overset{defn}{=}a_{0}$ and $W_{1}\left( a_{1},V\right) \overset{defn}{=}%
Q^{-1}\left( V,a_{1}\right) $. We will now show that for $y-p\in \Omega _{1}$
and correspondingly, $y\in \left[ L\left( y-p\right) ,U\left( y-p\right) %
\right] $, the functions $W_{0}\left( y,V\right) $ and $W_{1}\left(
y-p,V\right) $ will rationalize the choice-probabilities $q\left(
y,y-p\right) $.

To prove this, note that for any $v\in \lbrack 0,1]$, and $\left(
a_{0},a_{1}\right) \in \Omega $,%
\begin{equation}
a_{0}\leq Q^{-1}\left( v,a_{1}\right) \overset{\text{by }Q\left( \cdot
,a_{1}\right) \text{ }non\uparrow }{\Longrightarrow }Q\left(
a_{0},a_{1}\right) \geq \underset{=v\text{, by Lemma 3}}{\underbrace{Q\left(
Q^{-1}\left( v,a_{1}\right) ,a_{1}\right) }}\Longrightarrow Q\left(
a_{0},a_{1}\right) \geq v\text{.}  \label{a}
\end{equation}

Also, by definition of $Q^{-1}\left( \cdot ,a_{1}\right) $ as the supremum
in (\ref{4'}), we have that%
\begin{equation}
Q\left( a_{0},a_{1}\right) \geq v\Longrightarrow a_{0}\leq Q^{-1}\left(
v,a_{1}\right) \text{.}  \label{b}
\end{equation}%
Therefore, by (\ref{a}) and (\ref{b}), we have that $Q\left(
a_{0},a_{1}\right) \geq v\Longleftrightarrow a_{0}\leq Q^{-1}\left(
v,a_{1}\right) $. Thus, for $V\simeq U\left( 0,1\right) $, it follows that%
\begin{equation}
\Pr \left( Q^{-1}\left( V,a_{1}\right) \geq a_{0}\right) =\Pr \left( V\leq
Q\left( a_{0},a_{1}\right) \right) =Q\left( a_{0},a_{1}\right) \text{.}
\label{X'}
\end{equation}

Recall that for $y-p\in \Omega _{1}$ and correspondingly $y\in \left[
L\left( y-p\right) ,U\left( y-p\right) \right] $, we have that $Q\left(
y,y-p\right) =q\left( y,y-p\right) $ by definition. Therefore, it follows
from (\ref{X'}) that the utility functions $W_{0}\left( y,V\right) \equiv y$
and $W_{1}\left( y-p,V\right) \equiv Q^{-1}\left( V,y-p\right) $ with
heterogeneity $V\simeq Uniform\left( 0,1\right) $ rationalize the choice
probability function $q\left( \cdot ,\cdot \right) $ on its domain.

Next, note that $Q^{-1}\left( v,a_{1}^{\prime }\right) \leq Q^{-1}\left(
v,a_{1}\right) $ whenever $a_{1}^{\prime }<a_{1}$. To see this, suppose $%
a_{1}>a_{1}^{\prime }$ and yet $Q^{-1}\left( v,a_{1}\right) <Q^{-1}\left(
v,a_{1}^{\prime }\right) $. Choose $c$ s.t. $Q^{-1}\left( v,a_{1}\right)
<c<Q^{-1}\left( v,a_{1}^{\prime }\right) $. Then by conclusion (i) of the
previous lemma and by definition (\ref{4'}) of $Q^{-1}\left( v,\cdot \right) 
$, we must have $Q\left( c,a_{1}\right) <v\leq Q\left( c,a_{1}^{\prime
}\right) $. But since $a_{1}>a_{1}^{\prime }$, this contradicts conclusion
(1) of the Claim 1.

Next, it follows from (A) and (B) that $Q^{-1}\left( v,\cdot \right) $ is
continuous. To see this, fix $v\in \left[ 0,1\right] $, and suppose to the
contrary that $Q^{-1}\left( v,\cdot \right) $ is discontinuous at $a_{1}$;
suppose there exists $\epsilon >0$ such that for any $\delta >0$, $%
Q^{-1}\left( v,a_{1}\right) >Q^{-1}\left( v,a_{1}^{\prime }\right)
+\varepsilon $ for all $a_{1}^{\prime }$ satisfying $a_{1}^{\prime
}<a_{1}<a_{1}^{\prime }+\delta $. For any such $a_{1}^{\prime }$ satisfying $%
Q^{-1}\left( v,a_{1}\right) >Q^{-1}\left( v,a_{1}^{\prime }\right)
+\varepsilon $, it follows from the definition of $Q^{-1}\left( \cdot
,a_{1}^{\prime }\right) $ that there exists $\varepsilon ^{\prime
}=\varepsilon ^{\prime }\left( \varepsilon \right) >0$ s.t.%
\begin{equation}
Q\left( Q^{-1}\left( v,a_{1}\right) ,a_{1}^{\prime }\right) \overset{(1)}{%
\leq }Q\left( Q^{-1}\left( v,a_{1}^{\prime }\right) ,a_{1}^{\prime }\right)
-\varepsilon ^{\prime }\overset{\text{by Lemma 3}}{=}v-\varepsilon ^{\prime }%
\overset{\text{by Lemma 3}}{=}Q\left( Q^{-1}\left( v,a_{1}\right)
,a_{1}\right) -\varepsilon ^{\prime }\text{.}  \label{5'}
\end{equation}%
Inequality $(1)$ follows because $Q\left( Q^{-1}\left( v,a_{1}^{\prime
}\right) ,a_{1}^{\prime }\right) \leq Q\left( Q^{-1}\left( v,a_{1}\right)
,a_{1}^{\prime }\right) $ since $Q^{-1}\left( v,a_{1}\right) >$ $%
Q^{-1}\left( v,a_{1}^{\prime }\right) $, and if $Q\left( Q^{-1}\left(
v,a_{1}^{\prime }\right) ,a_{1}^{\prime }\right) =Q\left( Q^{-1}\left(
v,a_{1}\right) ,a_{1}^{\prime }\right) $ with $Q^{-1}\left( v,a_{1}\right) >$
$Q^{-1}\left( v,a_{1}^{\prime }\right) +\varepsilon $, then that contradicts
the definition of $Q^{-1}\left( v,a_{1}^{\prime }\right) $ as the sup.
Therefore, it follows from (\ref{5'}) that%
\begin{equation*}
Q\left( Q^{-1}\left( v,a_{1}\right) ,a_{1}\right) -Q\left( Q^{-1}\left(
v,a_{1}\right) ,a_{1}^{\prime }\right) \geq \varepsilon ^{\prime }\text{,}
\end{equation*}%
which contradicts that $Q\left( \cdot ,\cdot \right) $ is continuous in its
second argument for fixed value of its first argument (see property (2) in
Claim 1 above), since $a_{1}^{\prime }$ can be made arbitrarily close to $%
a_{1}$ by choosing $\delta $ small enough.

Finally, $W_{0}\left( y,\eta \right) =y$ is obviously continuous and
strictly increasing in $y$, thus (A') holds. Finally, (B) ensures that (B')
is satisfied.
\end{proof}

\begin{center}
\textbf{Reference}
\end{center}

1. Apostol, Tom M. (1974): Mathematical Analysis, Addison-Wesley.

2. Kruse, R.L. and Deely, J.J. 1969. Joint continuity of monotonic
functions. The American Mathematical Monthly, 76(1), pp.74-76.

3. Taylor, Angus E. (1955):\ Advanced Calculus, Ginn and Company.

\end{document}